\documentclass[aps,prd,superscriptaddress,onecolumn,showpacs,nofootinbib,preprintnumbers,notitlepage]{revtex4}
\usepackage{amsmath,amssymb}
\usepackage{hyperref}
\usepackage{hepunits}
\usepackage{graphicx,color}
\usepackage{mathtools}
\usepackage{todonotes}

\hypersetup{
    pdfnewwindow=true,      
    colorlinks=true,       
    linkcolor=blue,          
    citecolor=blue,        
    filecolor=blue,      
    urlcolor=blue        
}

\usepackage{xspace}
\usepackage{slashed}

\newcommand{\diff}{\textrm{d}}
\newcommand{\ket}[1]{\left|#1\right\rangle}
\newcommand{\bra}[1]{\left\langle#1\right|}
\newcommand{\braket}[2]{\left\langle#1\right.\left|#2\right\rangle}

\DeclareMathOperator{\re}{Re}
\DeclareMathOperator{\im}{Im}

\DeclareMathOperator{\const}{const}

\newcommand{\aka}{{\it a.k.a.}\xspace}
\newcommand{\eg}{{\it e.g.}\xspace}
\newcommand{\ie}{{\it i.e.}\xspace}

\newcommand{\bonn}{Universit\"at Bonn,
Helmholtz-Institut f\"ur Strahlen- und Kernphysik,
53115 Bonn, Germany}
\newcommand{\ceem}{Center for  Exploration  of  Energy  and  Matter,
Indiana  University,
Bloomington,  IN  47403,  USA}
\newcommand{\ghent}{Department of Physics and Astronomy,
Ghent University, Belgium}
\newcommand{\icn}{Instituto de Ciencias Nucleares,
Universidad Nacional Aut\'onoma de M\'exico,
Ciudad de M\'exico 04510, Mexico}
\newcommand{\indiana}{Physics  Department,
Indiana  University,
Bloomington,  IN  47405,  USA}
\newcommand{\jlab}{Theory Center,
Thomas  Jefferson  National  Accelerator  Facility,
Newport  News,  VA  23606,  USA}

\newcommand{\murcia}{Departamento de F\'isica,
Universidad de Murcia,
E-30071 Murcia, Spain}
\newcommand{\ucm}{Departamento de F\'isica Te\'orica,
Universidad Complutense de Madrid,
E-28040 Madrid, Spain}
\newcommand{\ectstar}{European Centre for Theoretical Studies in Nuclear Physics and Related
Areas (ECT$^*$) and Fondazione Bruno Kessler,
I-38123 Villazzano (TN), Italy}
\newcommand{\jpac}{Joint Physics Analysis Center}

\begin{document}

\title{Pole position of the $a_1(1260)$ from $\tau$-decay}

\author{M.~Mikhasenko}\email{mikhail.mikhasenko@hiskp.uni-bonn.de}\affiliation{\bonn}
\author{A.~Pilloni}\affiliation{\jlab}\affiliation{\ectstar}
\author{A.~Jackura}\affiliation{\ceem}\affiliation{\indiana}
\author{M.~Albaladejo}\affiliation{\jlab}\affiliation{\murcia}
\author{C.~Fern\'andez-Ram\'irez}\affiliation{\icn}
\author{V.~Mathieu}\affiliation{\jlab}
\author{J.~Nys}\affiliation{\ghent}
\author{A.~Rodas}\affiliation{\ucm}
\author{B.~Ketzer}\affiliation{\bonn}
\author{A.~P.~Szczepaniak}
\affiliation{\ceem}\affiliation{\indiana}\affiliation{\jlab}
\collaboration{\jpac}

\preprint{JLAB-THY-18-2816}


\newcommand{\lamnu}{\ensuremath{\lambda_\nu}}
\newcommand{\nut}{\ensuremath{\nu_\tau}}

\newcommand{\lami}[1]{\ensuremath{\lambda_{#1}}\xspace}
\newcommand{\lamO}{\lami{1}\xspace}
\newcommand{\lamT}{\lami{3}\xspace}
\newcommand{\lamsi}[1]{\ensuremath{\lambda_{s#1}}\xspace}
\newcommand{\lamsO}{\lamsi{1}\xspace}
\newcommand{\lamsT}{\lamsi{3}\xspace}

\newcommand{\sPTH}{\ensuremath{\sigma_\text{lim}}\xspace}
\newcommand{\sTH} {\ensuremath{\sigma_\text{th}}\xspace}
\newcommand{\sLIM}{\ensuremath{\sigma_\text{lim}}\xspace}

\newcommand{\rhopiS}{\ensuremath{\rho\pi\,S}}
\newcommand{\rhopiD}{\ensuremath{\rho\pi\,D}}
\newcommand{\ftwopiP}{\ensuremath{f_2\pi\,P}}

\newcommand{\rhoLAB}[1]{\ensuremath{\rho_{#1}}\xspace}
\newcommand{\rhoQTB}{\rhoLAB{\textrm{QTB}}\xspace}
\newcommand{\rhoSYMM}{\rhoLAB{\textrm{SYMM}}\xspace}
\newcommand{\rhoLABtilde}[1]{\ensuremath{\tilde{\rho}_{#1}}\xspace}
\newcommand{\rhoQTBtilde}{\rhoLABtilde{\textrm{QTB}}\xspace}
\newcommand{\rhoSYMMtilde}{\rhoLABtilde{\textrm{SYMM}}\xspace}
\newcommand{\rhoINT}{\rhoLAB{\textrm{INT}}\xspace}

\newcommand{\frho}{f_\rho}
\newcommand{\frhos}{\frho^*}
\newcommand{\frhoPQ}[1]{\frho^{(#1)}}
\newcommand{\frhoI}{\frhoPQ{I}}
\newcommand{\frhoII}{\frhoPQ{I\!I}}

\newcommand{\sr}{\sigma_\text{r}}

\newcommand{\tPQinv}[1]{t_{#1}^{-1}}
\newcommand{\tIinv}{\tPQinv{I}}
\newcommand{\tIIinv}{\tPQinv{I\!I}}

\newcommand{\MODEL}[1]{\ensuremath{\text{\tt #1}}}
\newcommand{\QTB}{\MODEL{QTB}\xspace}
\newcommand{\QTBDISP}{\MODEL{QTB-DISP}\xspace}
\newcommand{\SYMM}{\MODEL{SYMM}\xspace}
\newcommand{\DISP}{\MODEL{DISP}\xspace}
\newcommand{\SYMMDISP}{\MODEL{SYMM-DISP}\xspace}
\newcommand{\sQTBDISP}[1]{\text{\MODEL{sQTB-DISP}$^{(#1)}$}}

\newcommand{\approximateUnitarity}{approximate three-body unitary\xspace}
\newcommand{\threepion}{three-pion\xspace}
\newcommand{\lineshape}{line shape\xspace}
\newcommand{\lineshapes}{line shapes\xspace}
\newcommand{\parametrization}{parametrization\xspace}
\newcommand{\parametrizations}{\parametrization\ s\xspace}
\newcommand{\taudecay}{\ensuremath{\tau\to 3\pi\,\nu_\tau}\xspace}
\newcommand{\taudecaycharged}{\ensuremath{\tau^-\to \pi^-\pi^+\pi^-\,\nu_\tau}\xspace}
\newcommand{\ppiXp}{\ensuremath{\pi\,p\to3\pi\,p}\xspace}
\newcommand{\opp}{\ensuremath{J^{PC} = 1^{++}}\xspace}

\newcommand{\hattheta}{\hat{\theta}_{13}}

\newcommand{\XpolyOpt}[4]{\ensuremath{#4(#1,#2,#3)}\xspace}
\newcommand{\WpolyOpt}[3]{\XpolyOpt{#1}{#2}{#3}{W}}
\newcommand{\HpolyOpt}[3]{\XpolyOpt{#1}{#2}{#3}{H}}
\newcommand{\Wpoly}{\WpolyOpt{\sqrt{s}}{\sqrt{\sigma_1}}{\sqrt{\sigma_3}}}
\newcommand{\Hpoly}{\HpolyOpt{\sqrt{s}}{\sqrt{\sigma_1}}{\sqrt{\sigma_3}}}

\begin{abstract}
  We perform an analysis of the \threepion system with quantum numbers $J^{PC}=1^{++}$
  produced in the weak decay of $\tau$ leptons. The interaction is known to be
  dominated by the axial meson $a_1(1260)$.
  We build a model based on \approximateUnitarity
  and fix the free parameters by fitting it to the ALEPH data on \taudecaycharged decay.
  We then perform the analytic continuation of the amplitude to the complex energy plane.
  The singularity structures related to the $\pi\pi$ subchannel resonances are carefully addressed.
  Finally, we extract the $a_1(1260)$ pole position \mbox{$m_p^{(a_1(1260))}-i\Gamma_p^{(a_1(1260))}/2$} with
  \mbox{$m_p^{(a_1(1260))} = \unit{$(1209 \pm 4 ^{+12}_{-9})$}{\MeV}$},
  \mbox{$\Gamma_p^{(a_1(1260))} = \unit{$(576 \pm 11 ^{+89}_{-20})$}{\MeV}$}.
\end{abstract}

\maketitle

\section{Introduction}

The internal dynamics of the Quantum Chromodynamics (QCD) degrees of freedom manifests itself in the spectrum of hadron resonances.
The mass of a resonance characterizes the energy of the excitation while its width reflect on the coupling to the decay channels.
The meson spectrum has been qualitatively elucidated by the quark model~\cite{Tanabashi:2018oca} and recently, at least for some states,
calculations based on first principles lattice QCD  are becoming available~\cite{Durr:2008zz, Dudek:2010wm}.
For a majority of states, however, \textit{ab initio} QCD calculations of their decay properties, \eg decay widths,
branching ratios, are not yet available. Pushing such calculations forward is important given the growing body of evidence for novel hadronic phenomena
~\cite{Adolph:2015pws,Ketzer:2015tqa,Mikhasenko:2015vca,Aceti:2016yeb,Szczepaniak:2015hya,Szczepaniak:2015eza,Pilloni:2016obd},
\eg the $X,Y,Z$ states observed in heavy quarkonia~\cite{Esposito:2016noz,Guo:2017jvc,Olsen:2017bmm}.
Many of these new states are observed in decays to three-particle final sates.
While hadron scattering involving two stable particles is rather well understood formally,
the methodology for incorporating three and more particles is still being developed both
in the infinite volume~\cite{Aitchison:1966lpz,Holman:1965j,Fleming:1964zz,Eden:1966dnq,Mai:2017vot}
and finite volume~\cite{Hansen:2016ync,Briceno:2017tce,Briceno:2018mlh,Meng:2017jgx,Romero-Lopez:2018rcb,Mai:2017bge}.

A large number of light meson resonances dominantly decay to three pions.
This includes the enigmatic $a_1(1260)$ resonance, which is the lightest axial vector meson with $J^{PC}=1^{++}$.
The properties of the $a_1$ resonance are difficult to assess, due to its large width that is affected by the three-pion dynamics.
The $\pi\pi$ subchannel is dominated by the $\rho$ resonance whose finite width is expected to be important for the extraction of the $a_1$ resonance properties.
Indeed, a large part of the $a_1(1260)$ peak seen in the invariant mass distribution of three pions lays below the nominal
$\rho\pi$ threshold. However, the pole of the resonance was previously addressed in Lagrangian-based models~\cite{Roca:2005nm,Nagahiro:2011jn},
assuming a stable $\rho$-meson.

The \opp three-pion state can be observed in the \taudecay decay as well as in pion diffraction off a proton target \ppiXp.
There appears to be a discrepancy in the $a_1$ resonance parameters extracted from the two reactions~\cite{Tanabashi:2018oca,Eidelman:2003xyz}.
The problem
may be related to the presence of a large, coherent,
non-resonant background, known as the Deck process in pion diffraction~\cite{Ascoli:1974sp, Basdevant:1977ya,Deck:1964hm,Akhunzyanov:2018lqa}.
This process happens to dominate in the $J^{PC}=1^{++}$ partial wave and directly influences the extraction of
the $a_1(1260)$ resonance parameters in pion diffraction.
Thus, an independent determination of the $a_1(1260)$ resonance properties is not only relevant for a better understanding
of this state but also to constrain the Deck process,
which contributes significantly to other partial waves including the ones with the exotic quantum numbers $1^{-+}$~\cite{Akhunzyanov:2018lqa}.
In this paper, we therefore focus on the \taudecaycharged decay with the aim of extracting the  $a_1(1260)$ resonance parameters.

The paper is organized as follows. In Sec.~\ref{sec:reaction.model} we present our model, we relate the differential width of \taudecaycharged
to the three-pion scattering amplitude in the $1^{++}$ sector.
In Sec.~\ref{sec:fit} we show how the model is constrained by the fit to ALEPH data.
In Sec.~\ref{sec:poles} we explore the analytic properties of our model for
complex values of the $3\pi$ invariant mass squared, establishing the main singularities of the amplitude,
and we determine the location of the $a_1(1260)$ pole.
The studies of the systematics are described in Sec.~\ref{sec:systematic.studies}.
Our conclusions are summarized  in Sec.~\ref{sec:conclusion}.


\section{The reaction model}
\label{sec:reaction.model}

\begin{figure*}
  \includegraphics{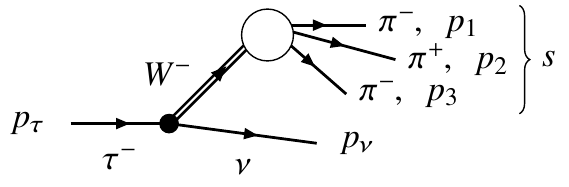}
  \caption{Diagram for the decay $\taudecaycharged$. The momenta of the $\tau$ lepton and $\nut$ are denoted by $p_\tau$ and $p_\nu$.
  The pions momenta are labeled by $p_i$, $i = 1,2,3$. $s$ is the invariant mass of the three pions.
  }
  \label{fig:tau}
\end{figure*}

We consider the reaction $\taudecay$ and derive an expression for the differential width which characterizes the $3\pi$ invariant mass spectrum~\cite{Bowler:1987bj,Kuhn:1992nz,Isgur:1988vm,Dumm:2009va,Nugent:2013hxa}.
The differential width is calculated by averaging (summing) over the $\tau$ ($\nut$) polarizations and integrating the matrix element squared over the final-state momenta,
\begin{equation} 
 \diff \Gamma = \frac{1}{2m_\tau} \cdot \frac{1}{2}\sum_{\lambda_\tau\lamnu} \left|A_{\lamnu,\lambda_\tau}\right|^2 \diff\Phi_4,
\end{equation}
where $m_\tau$ is the mass of the $\tau$-lepton, \mbox{$m_\tau = \unit{1776}{\MeV}$}~\cite{Tanabashi:2018oca},
the neutrino is considered massless, $\diff\Phi_4$ is the four-body differential phase space, and $\lambda_{x}$ are the lepton helicities of the \mbox{$x = \tau, \nu$}. The process is dominated by the emission of a $W$ boson by the leptonic current,
\begin{equation} 
  \bra{3\pi\,\nut,\lamnu}T\ket{\tau,\lambda_\tau} = -\frac{G_F}{\sqrt{2}} V^*_{ud} \,\bar{u}(p_\nu,\lamnu) \gamma^\alpha (1-\gamma^5) u(p_\tau,\lambda_\tau) \bra{3\pi } J^{5-}_\alpha (0) \ket{0},
\end{equation}
where $\bra{3\pi\,\nut,\lamnu}T\ket{\tau,\lambda_\tau} = A_{\lamnu,\lambda_\tau}\,(2\pi)^4\delta^4(p_\tau-p_\nu-p_{3\pi})$,
$G_F V^*_{ud}/\sqrt{2}$ is the Cabibbo-favored weak coupling,
$p_{3\pi}$, $p_\tau$, and $p_\nu$ are the four-momenta of three-pion system and the leptons, $u$ ($\bar{u}$) are the Dirac spinors of the $\tau$ ($\nu_\tau$), see Fig.~\ref{fig:tau}.
Because of $G$-parity conservation the $\pi^-\pi^+\pi^-$ system has positive $C$-parity.
Hence, the vector current $\bar u \gamma^\alpha u$ does not couple it, and can be removed.
Since the $W^-$ is heavily off-shell, one should also consider the timelike polarization, which carries $J^{PC} = 0^{-+}$.
However, the corresponding helicity amplitude is suppressed by the PCAC~\cite{Kuhn:1992nz,Colangelo:1996zp}.
This enables us to treat the off-shell $W^-$ as purely axial. The polarization of the real $W^-$ provides a complete basis which we use to expand the hadronic current,
\begin{equation} 
  A_{\lamnu,\lambda_\tau} = \frac{G_F}{\sqrt{2}} V^*_{ud} \,\bar{u}(p_\nu,\lamnu) \gamma^\alpha \gamma^5 u(p_\tau,\lambda_\tau)  \sum_{\Lambda} \varepsilon_\alpha(\Lambda) A_\Lambda,
\end{equation}
where $\varepsilon^{\alpha*}(\Lambda)\bra{ 3\pi } J^{5-}_\alpha (0) \ket{0} = A_{\Lambda}\,(2\pi)^4\delta^4(p_\tau-p_\nu-p_{3\pi})$ is the helicity amplitude
for the decay of the axial current to three pions.
The squared matrix element summed and averaged over the $\nu_\tau$ and $\tau$ helicities, respectively, is
\begin{equation} \label{eq:contraction} 
  \frac{1}{2}\sum_{\lambda_\tau\lamnu}  \left|A_{\lamnu,\lambda_\tau} \right|^2 = G_F^2|V_{ud}|^2\,
  \left(p_\tau^\alpha p_\nu^\beta + p_\tau^\beta p_\nu^\alpha- g^{\alpha\beta}(p_\tau\cdot p_\nu)\right)\sum_{\Lambda,\Lambda'} \varepsilon_\alpha(\Lambda) \varepsilon_\beta^*(\Lambda') \,A_{\Lambda}A_{\Lambda'}^*.
\end{equation}
The explicit evaluation of the expression is performed in the $\tau$-rest frame where
$p_\tau\cdot \varepsilon(0) = (m_\tau^2-s)/(2\sqrt{s})$, and $p_\tau\cdot \varepsilon(\pm) = 0$.

Using the recursive relation for the phase space, we split it into
the $\tau^-\to W^-\nut$-phase space $\diff \Phi_2$, and the \threepion phase space $\diff\Phi_3$: $\diff\Phi_4 = \int \diff\Phi_2 \, \diff \Phi_3\,\diff s /(2\pi)$, where $\sqrt{s}$ is the invariant mass of the hadronic system.
To obtain the differential width $\diff\Gamma/\diff s$, we integrate explicitly over the neutrino angles,
\begin{equation}\label{eq:diff.width} 
  \frac{\diff \Gamma}{\diff s} = \frac{G_F^2 \left|V_{ud}\right|^2}{64\pi^2\,m_\tau^3}\,
  (m_\tau^2-s)^2
  \int \diff \Phi_3
  \left(|A_+|^2+|A_-|^2+\frac{m_\tau^2}{s}|A_0|^2\right).
\end{equation}
Here, one power of the factor $(m_\tau^2-s)$ follows from the matrix element in Eq.~\eqref{eq:contraction},
the other is given by the $W^-\nut$ two-body phase space.
The expression for the $\diff \Phi_3$ is given in Appendix~\ref{app:integrals}.
The integral is kept in the final expression
to facilitate the further discussion on partial-wave expansion of the amplitude $A_\Lambda$.

The helicity amplitude $A_\Lambda$ describes the coupling of the axial current to the three charged pions.
The pions are labeled as follows, $\pi_1^-\pi_2^+\pi_3^-$ (see Fig.~\ref{fig:tau}).
We use the isobar model to parametrize the dynamics and explicitly incorporate the $\pi_1^-\pi_3^-$ Bose symmetry, 
\begin{equation}\label{eq:isospin} 
  A_{\Lambda} = A_{\Lambda}^{(3)}+A_{\Lambda}^{(1)},
\end{equation}
where the isobar amplitude $A_{\Lambda}^{(k)}$ includes only the subchannel interaction in a pion pair leaving the pion indexed $k$ as a bachelor.
In Eq.~\eqref{eq:isospin}, we disregard the $\pi^-\pi^-$ interaction since it is negligible compared to the dominant $\rho$-meson in the $\pi^+\pi^-$ subchannel.
The pion momenta are denoted by $p_i$ where $i = 1,2,3$ as shown in Fig.~\ref{fig:tau} and the subchannel invariant mass squared is denoted as $\sigma_k = (p_i+p_j)^2$.
Here and below we use the circular convention, \ie the bachelor pion has index $k$ such that the $(ijk)$ are numbers $(123)$, $(231)$ or $(312)$.

Each isobar amplitude receives different contributions, often referred to as decay channels~\cite{Tanabashi:2018oca}.
The importance of different decay channels can be estimated by the relative branching fractions of the $a_1(1260)$ decay.
The latest measurements were carried out by the CLEO experiment from $\tau$ decay~\cite{Asner:1999kj,Briere:2003fr} and by the COMPASS experiment in diffractive production ~\cite{Adolph:2015tqa}.
The extraction of branching ratios is model-dependent and is influenced by the production mechanism; however, we get a rough estimate of their relative importance.
The \rhopiS-wave channel is dominant with a branching ratio of $60\%-80\%$.
The second most important channel, $f_0(500)\pi\,P$-wave, was estimated to contribute less than $20\%$.
The combined branching ratio to the remaining channels (\rhopiD-wave, \ftwopiP-wave, $K^{*} \bar K\,S,D$-waves) does not exceed $10\%$.
We thus limit the analysis to the main \rhopiS-wave channel.
Including other decay channels would require the introduction of additional parameters for couplings and production strengths, which cannot be fixed by current publicly available data.

Therefore, we take the isobar amplitude to have the form,
\begin{equation} \label{eq:partial.wave.decomposition} 
  A_{\Lambda}^{(k)} = \mathcal{C}^{(k)} \, a(s) \frho(\sigma_k)N_{\Lambda}(\Omega_k,\Omega_{ij}),
\end{equation}
where $\mathcal{C}^{(k)} = \braket{1,\mu_i;1,\mu_j}{1,0} = \pm 1/\sqrt{2}$ is the Clebsch-Gordan coefficient relating
the two pion with isospin projection $\mu_{i,j}=\pm 1$ to $\rho^{0}$ isospin states, thus, the sign depends on the index $k$.
The $a(s)$ denotes the dynamical part of the amplitude $a_1\to\rho\pi\,S$-wave in the canonical basis~\cite{Collins:1977xx, Martin:1970xx},
$\frho(\sigma)$ is a \parametrization for the $\rho$-meson decay amplitude, and
$N_\Lambda(\Omega_k,\Omega_{ij})$ is the angular decay function for the decay chain $a_1 \to \rho\pi$, $\rho\to \pi\pi$,
\begin{equation} \label{eq:decay.function} 
  N_\Lambda(\Omega_k,\Omega_{ij}) = \sqrt{3} \sum_\lambda D^{1*}_{\Lambda\lambda}(\Omega_k) D^{1*}_{\lambda0}(\Omega_{ij})~.
\end{equation}
The angles $\Omega_k = (\theta_k,\phi_k)$ are the polar and the azimuthal angles of the
vector $\vec{p}_i+\vec{p}_j$ in the $3\pi$ helicity frame, i.e. the center-of-mass (CM) frame with the axis orientation fixed by the production kinematics.
The $\Omega_{ij} = (\theta_{ij},\phi_{ij})$ are the spherical angles of the pion $i$ in the helicity frame of the isobar $(ij)$.
Detailed discussion on the decay function in Eq.~\eqref{eq:decay.function} can be found in the Appendix~\ref{app:integrals}.

The \lineshape of the $\rho$-meson is given by the customary Breit-Wigner amplitude with dynamical width~\cite{Pisut:1968zza, Adolph:2015tqa}
\begin{equation}\label{eq:isobar.parametrization} 
  \frho(\sigma) = \mathcal{N} \frac{F_1\left(p(\sigma)R\right)}{m_\rho^2-\sigma-im_\rho \Gamma_\rho(\sigma)},\qquad
  \Gamma_\rho(\sigma) = \Gamma_\rho \frac{p(\sigma)}{p(m_\rho^2)} \frac{m_\rho}{\sqrt{\sigma}}  \frac{F_1^2\left(p(\sigma)R\right)}{F_1^2\left(p(m_\rho^2)R\right)},\qquad
  F_1^2(pR) = \frac{(pR)^2}{1+(pR)^2},
\end{equation}
where $p(\sigma) = \sqrt{\sigma/4-m_\pi^2}$ is the pion break-up momentum,
the function $F_1(pR)$ combines the threshold factor $p(\sigma)$ and the customary Blatt-Weisskopf barrier factor with size parameter $R = 5\,\GeV^{-1}$. We use in the analysis
 \mbox{$m_\pi = \unit{139.57}{\MeV}$}, \mbox{$m_\rho = \unit{775.26}{\MeV}$}~\cite{Tanabashi:2018oca}.
For convenience we fix $\mathcal{N}$ so that the phase-space integral $\rho(s)$ defined below in Eq.~\eqref{eq:rho} approaches the two-body phase space
asymptotic value, $1/8\pi$, in the limit $s\to\infty$, \ie
 \begin{equation} \label{eq:frho.norm} 
   \int_{4m_\pi^2}^{\infty} \sqrt{1-4m_\pi^2/\sigma}\,|\frho(\sigma)|^2 \diff \sigma = 16\pi^2.
 \end{equation}
 The normalization for $\frho(\sigma)$ fixes the normalization of $a(s)$.

\begin{figure*}
  \includegraphics[width=0.45\textwidth]{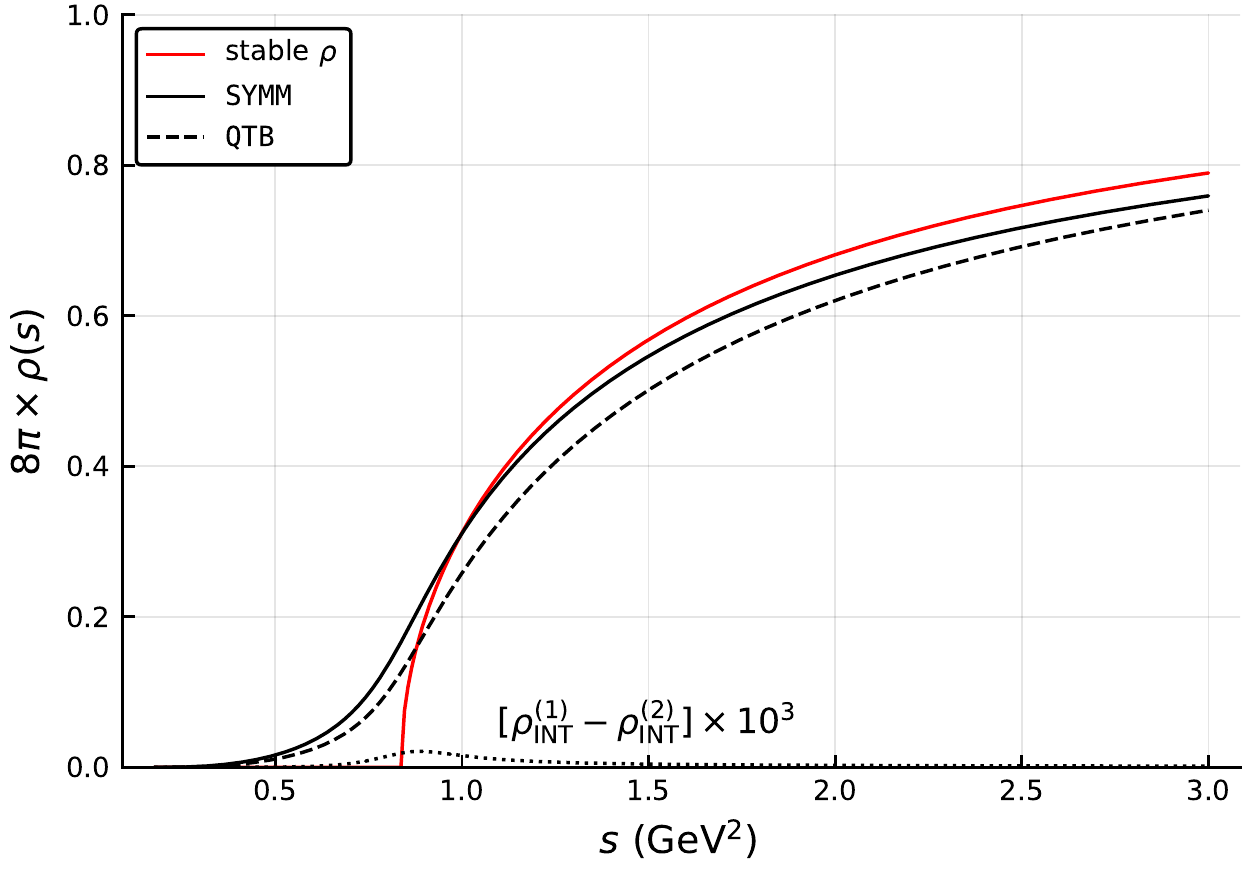}
  \caption{The phase space $\rho(s)$ calculated for different models. The black solid line shows the symmetrized $\rhoSYMM$ from Eq.~\eqref{eq:rho}.
  The dashed curve represents \rhoQTB from Eq.~\eqref{eq:rho.qtb}, which neglects the interference between the two $\rho\pi$ decay chains. For reference we draw the two-body $\rho\pi$ phase space given by
  $\sqrt{(s-(m_\rho+m_\pi)^2)(s-(m_\rho-m_\pi)^2)}/(8\pi s)$ with a solid red line.
  Due to the chosen normalization in Eq.~\eqref{eq:frho.norm}, all functions approach the same asymptotic limit.
  The dotted line shows the difference in the interference terms calculated in two different ways for $s+i\epsilon$ as discussed in Sec.~\ref{sec:symm}.
  }
  \label{fig:phase.space}
\end{figure*}

Using Eqs.~\eqref{eq:isospin},\eqref{eq:partial.wave.decomposition} to substitute the amplitude $A_\Lambda$ in Eq.~\eqref{eq:diff.width},
we get the expression for the differential width in terms of the dynamic amplitude $a(s)$.
\begin{equation}\label{eq:diff.width.through.a} 
  \frac{\diff \Gamma}{\diff s} = \frac{G_F^2 \left|V_{ud}\right|^2}{64\pi^2\,m_\tau^3}\,
  \left(2+\frac{m_\tau^2}{s}\right)(m_\tau^2-s)^2 \,|a(s)|^2 \rho(s).
\end{equation}
where $\rho(s)$ is the effective $\rho\pi$ phase space. We will consider two models for $\rho(s)$'s:
\begin{subequations}
  \begin{align}\label{eq:rho} 
    \rhoSYMM(s) &= \frac{1}{2}\int\diff\Phi_3 \left|\frho(\sigma_1) N_{0}(\Omega_1,\Omega_{23}) - \frho(\sigma_3)N_{0}(\Omega_3,\Omega_{12})\right|^2,\\
    \label{eq:rho.qtb} 
    \rhoQTB(s) &= \int \diff \Phi_3 |\frho(\sigma_1) N_0(\Omega_1,\Omega_{23})|^2.
  \end{align}
\end{subequations}
The expression in Eq.~\eqref{eq:rho} strictly follows from Eqs.~\eqref{eq:isospin}, \eqref{eq:partial.wave.decomposition}, and \eqref{eq:diff.width.through.a}.
The label $\SYMM$ is introduced to emphasize the symmetrization between the decay channels, \ie the $\rho\pi$ channels $k = 1$ and $3$.
The relative minus sign comes from the symmetry of the isospin coefficient in Eq.~\eqref{eq:partial.wave.decomposition}.
The integral in Eq.~\eqref{eq:rho} is the same for all helicities $\Lambda$ due to the properties of the Wigner $d$-functions, therefore we set $\Lambda=0$ for simplicity.
The interference term is only significant at low energy, where the overlapping region of the two $\rho$-mesons contributes to a substantial fraction of the Dalitz plot.
The $\rhoQTB$ (Quasi-Two-Body) in Eq.~\eqref{eq:rho.qtb} is a simplified phase space where the interference term is neglected.
In this case, the integrals of the two decay chains squared are identical, which cancels the $1/2$ factor in front.
This model treats the $\rho$-meson as quasi-stable and the interaction between the $\rho\pi$ as a two-body interaction.
The simplification is suggested and discussed in Ref.~\cite{Basdevant:1978tx} to treat the multiparticle final states.
The same approximation is commonly used to account for $4\pi$ channel in the $\pi\pi/K\bar{K}$ coupled-channels problem~\cite{Bugg:1996ki,Anisovich:2011zz}).
Finally, as shown in Fig.~\ref{fig:phase.space}, the interference is rather small.
Since this model is simpler, we would like to test it as an alternative.

Our model for the decay amplitude is constrained by the \approximateUnitarity~\cite{Basdevant:1978tx, Basdevant:1977ya}.
Turning general 3-body unitarity into some practically useful equations is cumbersome and not complete yet.
A significant progress in this direction has been made in Refs.~\cite{Aitchison:1966lpz,Fleming:1964zz,Eden:1966dnq,Holman:1965j}.
In particular, one can separate the genuine three-body unitary from the subchannel unitarity
related to rescattering between different isobars. 
These processes modify the \lineshape of the subchannel amplitudes~\cite{Aitchison:1979fj, Niecknig:2012sj, Niecknig:2015ija, Danilkin:2014cra, Albaladejo:2017hhj,
Colangelo:2016jmc, Lorenz:2017svg}.
A good example is the $\rho\pi$-dynamics studied in the $1^{--}$ sector in the decay of $\omega/\phi$~\cite{Niecknig:2012sj, Danilkin:2014cra},
where the final-state interaction were found to shift and skew the $\rho$-meson peak.
Conversely, in our models we focus on the 3-body resonance dynamics, and simplify the problem by neglecting the effects of the rescattering on the isobar \lineshapes.
We introduce the $\rho\pi$ elastic scattering isobar amplitude $t(s)$, to impose the unitarity constraints for the amplitude $a(s)$:
\begin{subequations}
  \label{eq:unitarity}
\begin{align} \label{eq:unitarity.a}  
  2\im \, a(s) &= t^*(s) \,\rho(s)\, a(s),\\ \label{eq:unitarity.t}
  2\im \, t(s) &= t^*(s) \,\rho(s)\, t(s),
\end{align}
\end{subequations}
where $\rho(s)$ is the effective phase space given by Eq.~\eqref{eq:rho} or Eq.~\eqref{eq:rho.qtb}.
The factor of $2$ in the left-hand-side of Eq.~\eqref{eq:unitarity} is kept for convenience.

The unitarity equations~\eqref{eq:unitarity} can be satisfied by a certain choice of the \parametrization. 
\begin{equation}\label{eq:amp} 
	t(s) = \frac{g^2}{m^2-s-i g^2 C(s)/2},\quad a(s) = \alpha(s)t(s),
\end{equation}
where $C(s)$ is an analytic function constrained by condition $\im \,iC(s) = \rho(s)$.
To describe the amplitude dominated by a single resonance, we added a first order polynomial $(m^2-s)/g^2$ to the denominator of $t(s)$, which is equivalent to have the $K$-matrix with a single pole~\cite{Martin:1970xx}.
The numerator function $\alpha(s)$ is supposed to incorporate
the singularities specific to the production process into the amplitude $a(s)$.
The final state interaction required by unitarity is accounted for by the multiplicative form of the production amplitude in Eq.~\eqref{eq:amp}. It diminishes the differences between different possible production mechanisms, \eg resonant vs. non-resonant production of $\rho\pi$.
In the case at hand we use $\alpha = \const$.
There are two common constructions for $C(s)$ which both satisfy unitarity:
\begin{enumerate}
  \item The models with $C(s) = \rho(s)$ will be called \textit{non-dispersive}.
Twhese models have left-hand singularities on the physical sheet inherited from the phase space, which are not motivated by physics.
  \item The \textit{dispersive} models have $C(s) = \tilde{\rho}(s)$, with
  \begin{equation}\label{eq:rho.tilde} 
    i\tilde{\rho}(s) = l_0 + \frac{s}{\pi}\int_{9m_\pi^2}^{\infty}	\diff s' \frac{\rho(s')}{s'(s'-s-i\epsilon)},
  \end{equation}
  where the subtraction constant $l_0$ is chosen such that the real part of $i\tilde{\rho}(s)$ is zero at the point $(m_\rho + m_\pi)^2$.
  The function $i\tilde{\rho}(s)$ has no singularities other than the unitarity cut as guaranteed by the Cauchy integral theorem. It is analogous to the Chew-Mandelstam function for the two-body scattering amplitude~\cite{Basdevant:1977ya}.
\end{enumerate}
We note that the first construction with $C(s) = \rho(s)$ resembles the Breit-Wigner amplitude with a dynamical width~\cite{Tanabashi:2018oca}. In contrast, the dispersive amplitudes do not have the unmotivated left-hand cut  generated by $\rho$ in Eq.~\eqref{eq:rho}.
For all models, the structure of $C(s)$ ensures unitarity and extends the applicability of Eq.~\eqref{eq:amp} from threshold to energy regions where higher-lying resonances or/and non-elastic channels become significant.

To summarize, the final expression for the differential cross section is.
\begin{equation}\label{eq:final.diff.width} 
	\frac{\diff \Gamma}{\diff s} = \frac{1}{s}\left(1-\frac{s}{m_\tau^2}\right)^2\left(1+\frac{2s}{m_\tau^2}\right) \frac{c \rho(s)}{\left|m^2-s-ig^2 C(s)/2\right|^2}.
\end{equation}
Eq.~\eqref{eq:final.diff.width} follows from Eq.~\eqref{eq:diff.width.through.a}. 
The constant $c$ absorbs all energy-independent numerical factors; $m$, $g$, and $c$ are real parameters
which are fitted to data.
The four models we are going to test are summarized in Table~\ref{tab:models}.
\begin{table*}
  \caption{Summary of the models discussed in Sec.~\ref{sec:reaction.model}.
  The numerator and denominator refer to Eq.~\eqref{eq:final.diff.width}.
  }
  \label{tab:models}
  \begin{ruledtabular}
  \begin{tabular}{c || c | c || r || c |c }
    Model  & $\rho(s)$ in the numerator & $C(s)$ in the denominator &
            $\chi^2/\text{n.d.f.}$ & $m$ (\GeV) & $g$ (\GeV) \\\hline
    \SYMMDISP  & $\rhoSYMM(s)$ & $\rhoSYMMtilde(s)$  &   $94/100$ & $1.205$ & $6.64$ \\ 
    \SYMM      & $\rhoSYMM(s)$ & $\rhoSYMM(s)$       &  $663/100$ & $1.230$ & $6.65$ \\
    \QTBDISP   & $\rhoQTB(s)$  & $\rhoQTBtilde(s)$   &   $68/100$ & $1.223$ & $7.45$ \\
    \QTB       & $\rhoQTB(s)$  & $\rhoQTB(s)$        &  $344/100$ & $1.236$ & $7.42$
  \end{tabular}
  \end{ruledtabular}
\end{table*}
Our primary model is \SYMMDISP, which is the one that incorporates the most of physical arguments.
The \SYMM model contains additional left-hand singularities with respect to \SYMMDISP.
The \QTB and \QTBDISP models do not include the interference between the two decay chains,
but are much simpler to calculate on the real axis and continue to the complex plane.
The $C(s)$ is calculated using the same $\rho(s)$ as in the numerator of Eq.~\eqref{eq:final.diff.width}, which is either $\rhoQTB$ or $\rhoSYMM$
as given in Table~\ref{tab:models}.


\section{Fit results and resonance parameters}
\label{sec:fit}
The largest public dataset for \taudecay was collected by the ALEPH experiment in 2005~\cite{Schael:2005am}.\footnote{
An updated analysis was published in 2014~\cite{Davier:2013sfa}.
The main difference is related to the use of a new method to unfold detector effects from the mass spectra. However,
the data were binned into wider bins with variable bin size, which makes it less straightforward to use. For this reason we stick to data of~\cite{Schael:2005am}.}
The distribution $\diff\Gamma/\diff s$ is binned in $0.025\,\GeV^2$ bins and normalized by the measured branching ratio.
We fit $103$ data points in the range $0.38\,\GeV^2 \le s \le 2.94\,\GeV^2$.
We minimize the $\chi^2$-function taking into account the covariance matrix provided in Ref.~\cite{Schael:2005am},
\begin{equation} 
	\chi^2(c,m,g) = (\vec{D} - \vec{M}(c,m,g))^T C_{\text{stat}}^{-1} (\vec{D} - \vec{M}(c,m,g)),
\end{equation}
where $\vec{D}$ is a vector of ALEPH data points,
$\vec{M}(c,m,g)$ is a vector of the model predictions calculated for the centers of the bins.
The matrix $C_{\text{stat}}$ is the covariance matrix of the statistical errors.
The systematic uncertainties are smaller than the statistical ones by a factor $5$, and can be neglected.
Nonzero correlations among different bins are introduced by the unfolding procedure.
It is worth noticing the $3\pi$ spectrum does not show the expected random noise.
As discussed in the follow up analysis of the ALEPH~\cite{Davier:2013sfa},
the problem appears because the errors of the unfolding procedure were not correctly propagated.
Hence, the absolute value of $\chi^2$ we obtained does not have a strict statistical meaning.
However, we assume that for the model characterization based on relative $\chi^2$ values,
the problem should not be critical.

\begin{figure*}
  \includegraphics[width=0.48\textwidth]{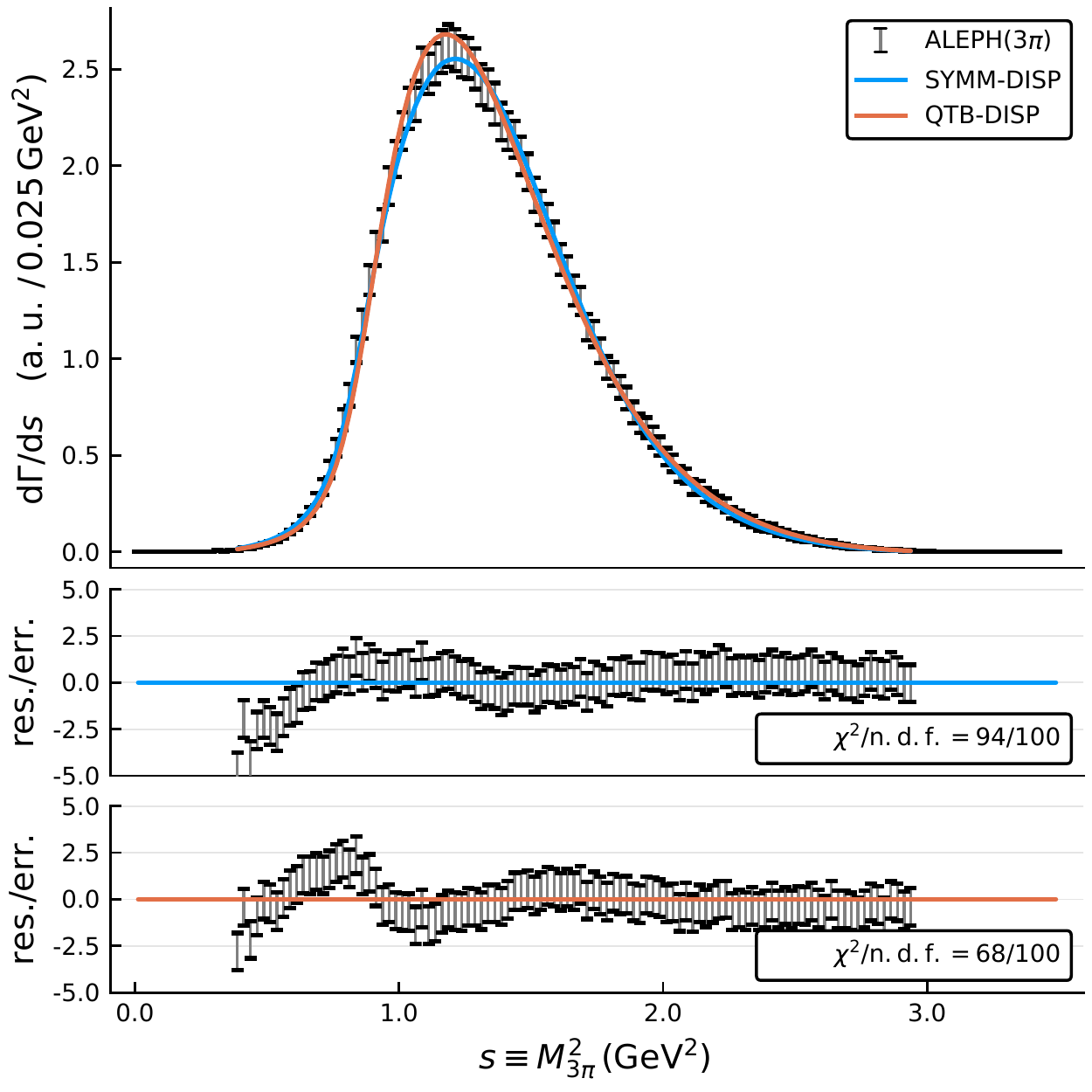}\qquad
  \includegraphics[width=0.48\textwidth]{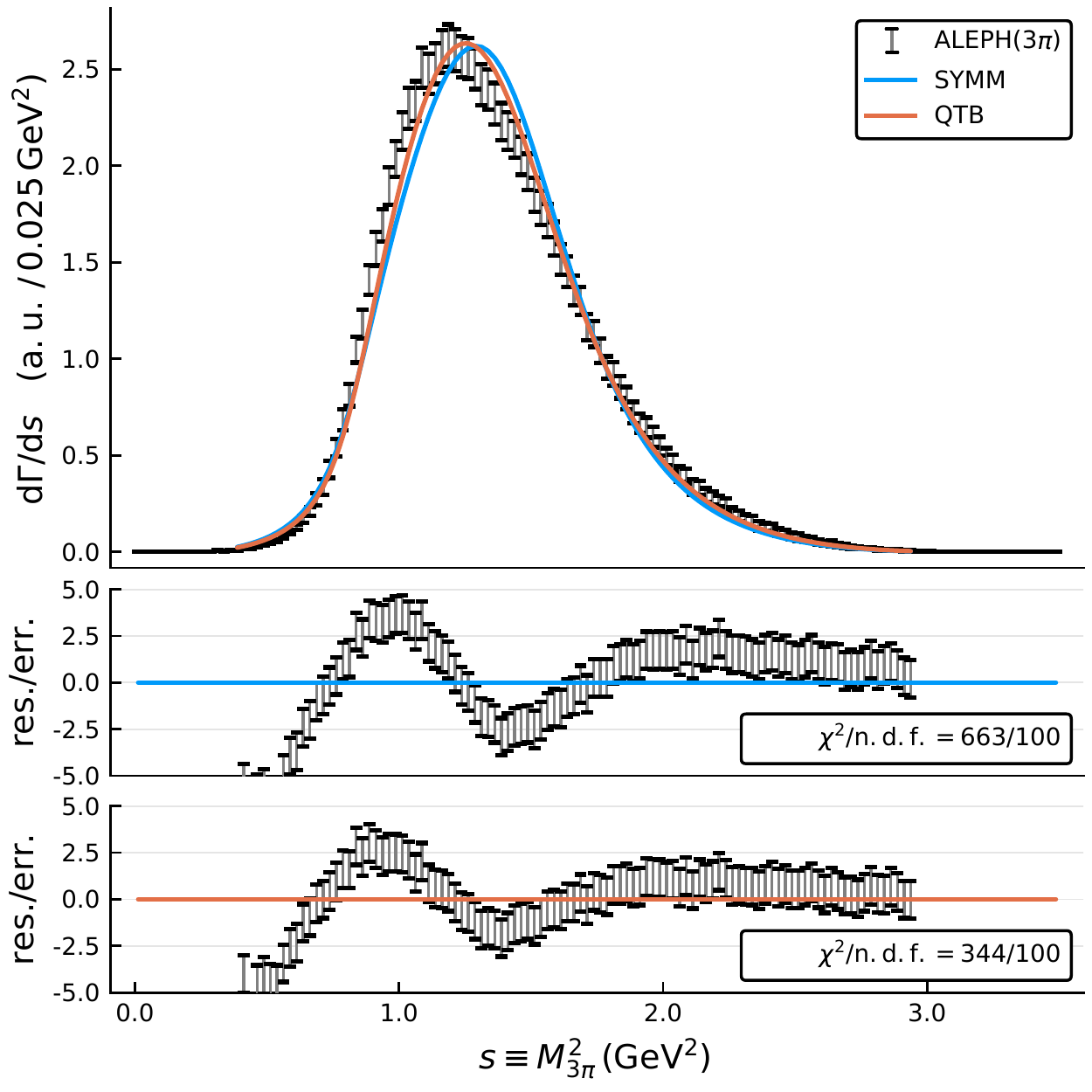}
  \caption{
  Fit to ALEPH data with the four  models described in the text.
  The models differ by either including the effect of interference between two $\rho\pi$ decay channels ({\SYMM}) or not ({\QTB}),
  and either using the dispersive integral over the phase space ({\DISP}), or not. The lower panels show the normalized residues.
  }
  \label{fig:fits}
\end{figure*}

The gradient minimization is performed using the {\tt NLopt} optimizer and
the {\tt ND\_MMA} algorithm~\cite{Johnson2011,Svanberg:2002xyz} with the automatic differentiation provided by the {\tt ForwardDiff.jl}-package~\cite{RevelsLubinPapamarkou2016}.
The minimum we find is always stable and isolated, as checked by repeating the minimization from different starting values.
Fits to the ALEPH dataset are shown in Figs.~\ref{fig:fits}, and the fit parameters and $\chi^2$ values are shown in Table~\ref{tab:models}.
The non-dispersive models are not consistent with the data, with $\chi^2$ at least three times worse than we have obtained for the dispersive models.
In particular, they fail to reproduce the \lineshape around the peak and in the threshold region,
and we do not consider them any further.
On the other hand, the dispersive models show a good agreement with data, obtaining
$\chi^2/\textrm{n.d.f.}=94/100$ and $\chi^2/\textrm{n.d.f.}=61/100$ for the
\SYMMDISP and \QTBDISP, respectively.

In the next section we will perform the analytic continuation of the amplitude to the second sheet and
search for the $a_1(1260)$ resonance pole.
For comparison with the PDG~\cite{Tanabashi:2018oca},
we first provide the customary Breit-Wigner parameters, that can be extracted on the real axis.
We remind the reader that these are expected to be reaction-dependent,
and do not provide an unambiguous characterization of the resonance.
We define the Breit-Wigner mass squared $m_{\text{BW}}^2$ as the value of $s$ when
the denominator of the amplitude $t(s)$ in Eq.~\eqref{eq:amp} becomes purely imaginary.
The value of the denominator at this point gives the Breit-Wigner width,
as it is equal to $-i m_{\text{BW}}\Gamma_{\text{BW}}$.
For \QTBDISP we get the Breit-Wigner mass and width as $(1246 \pm 3)\,\MeV$ and $(394 \pm 5)\,\MeV$;
for \SYMMDISP, $(1254 \pm 3)\,\MeV$ and $(461 \pm 8)\,\MeV$,
where the errors are statistical only.

\section{Analytic continuation the pole position}
\label{sec:poles}
Once the amplitude is fixed on the real axis, its analytic structure is unambiguously defined and can be explored.
Unitarity introduces a branch cut along the real axis from the $3\pi$ threshold to infinity, which opens a non-trivial Riemann topology or sheet structure.
The first Riemann sheet is the one containing the physical values of the amplitude slightly above the real axis.
By construction, the amplitudes in the dispersive models
contain no other singularity on the first sheet than the unitarity cut.
Resonance poles are expected to lie on the second sheet, which is connected to the physical axis from below.
The unitarity condition Eq.~\eqref{eq:unitarity.t} gives us a relation on the real axis that can be used to continue the amplitude in the complex $s$-plane.
The real-axis relation followed from Eq.~\eqref{eq:unitarity} reads
\begin{equation} \label{eq:tII-1=} 
  \tIinv(s + i\epsilon) - \Delta t^{-1}(s) = \tIinv(s- i\epsilon) = \tIIinv(s+ i\epsilon),
\end{equation}
where $\Delta t^{-1}(s)\equiv\tIinv(s+i\epsilon)-\tIinv(s-i\epsilon) = -i\rho(s)$ is the discontinuity across the cut,
$s$ is real, $\epsilon$ is an infinitesimal positive number, and the Roman subscript indicates the Riemann sheet.
Thus, $\tIIinv(s) = \tIinv(s)+i \rho(s)$ and the pole positions are determined by $\tIIinv(s) = 0$.
The first sheet amplitude, $\tIinv(s)$, is straightforward to calculate in the complex plane using the dispersive integral in Eq.~\eqref{eq:rho.tilde}.
Continuation of the discontinuity, however, is more challenging since it is not explicitly analytical expression,
as Eq.~\eqref{eq:rho} contains a modulus operator.
Therefore, we need to find an analytic function which coincides with the discontinuity on the real axis.
All singularities of the discontinuity $-i\rho(s)$ will be present in the second sheet amplitude according to Eq.~\eqref{eq:tII-1=}.
Among those, we expect the reflection of the $\rho\pi$ unitarity cut, which is pushed into the second sheet due to the unstable nature of the $\rho$-meson.

For the continuation to the complex $s$-plane, we need to evaluate $\frho(\sigma)$ and $\frhos(\sigma)$ in Eq.~\eqref{eq:rho} and Eq.~\eqref{eq:rho.qtb} for complex argument $\sigma$.
Along the physical axis $\frho(\sigma)=\frhoI(\sigma+i\epsilon)$ and
the analytic function $\frhoII(\sigma+i\epsilon)$ coincides with $\frhos(\sigma)$
due to the Schwarz reflection principle and the continuity of the Riemann sheet structure, since
\begin{equation}\label{eq:fstar} 
  \frhos(\sigma) = \frho^{(I)*}(\sigma+i\epsilon) = \frhoI(\sigma-i\epsilon) = \frhoII(\sigma+i\epsilon).
\end{equation}

\subsection{Analytic continuation of the \QTBDISP model}
\label{sec:qtb-cont}
We start with the \QTBDISP model, whose analytic continuation is simpler than the one of the \SYMMDISP model.
The discontinuity across the unitarity cut is given by $-i\rhoQTB$ in Eq.~\eqref{eq:rho.qtb}.
The angular integrals in the phase space can be solved analytically due to the properties of the Wigner $D$-functions. We obtain
\begin{equation} \label{eq:rho.qtb.complex} 
  \rhoQTB(s) = 
  \frac{1}{2\pi(8\pi)^2 s}\int_{4m_\pi^2}^{(\sqrt{s}-m_\pi)^2}
  \frhoII(\sigma_1)\frhoI(\sigma_1)
  \frac{\sqrt{\lamO\lamsO}}{\sigma_1}\, \diff \sigma_1,
\end{equation}
where we used the definition
$\lambda_i = \lambda(\sigma_i,m_\pi^2,m_\pi^2)$, $\lambda_{si} = \lambda(s,\sigma_i,m_\pi^2)$, with $\lambda$ being the K\"all\'en triangle function $\lambda(x,y,z) = x^2 + y^2 + z^2 - 2(xy + yz + zx)$.
Using Eq.~\eqref{eq:fstar}, we replaced $\left|\frho(\sigma_1)\right|^2$ by the analytic expression $\frhoII(\sigma_1)\frhoI(\sigma_1)$.
The function $\frhoI(\sigma_1)$ does not have singularities apart form cuts on the real axis,
while the $\frhoII(\sigma_1)$ contains the pole of the $\rho$-meson in the complex plane.
\begin{figure*}
  \includegraphics[width=0.45\textwidth]{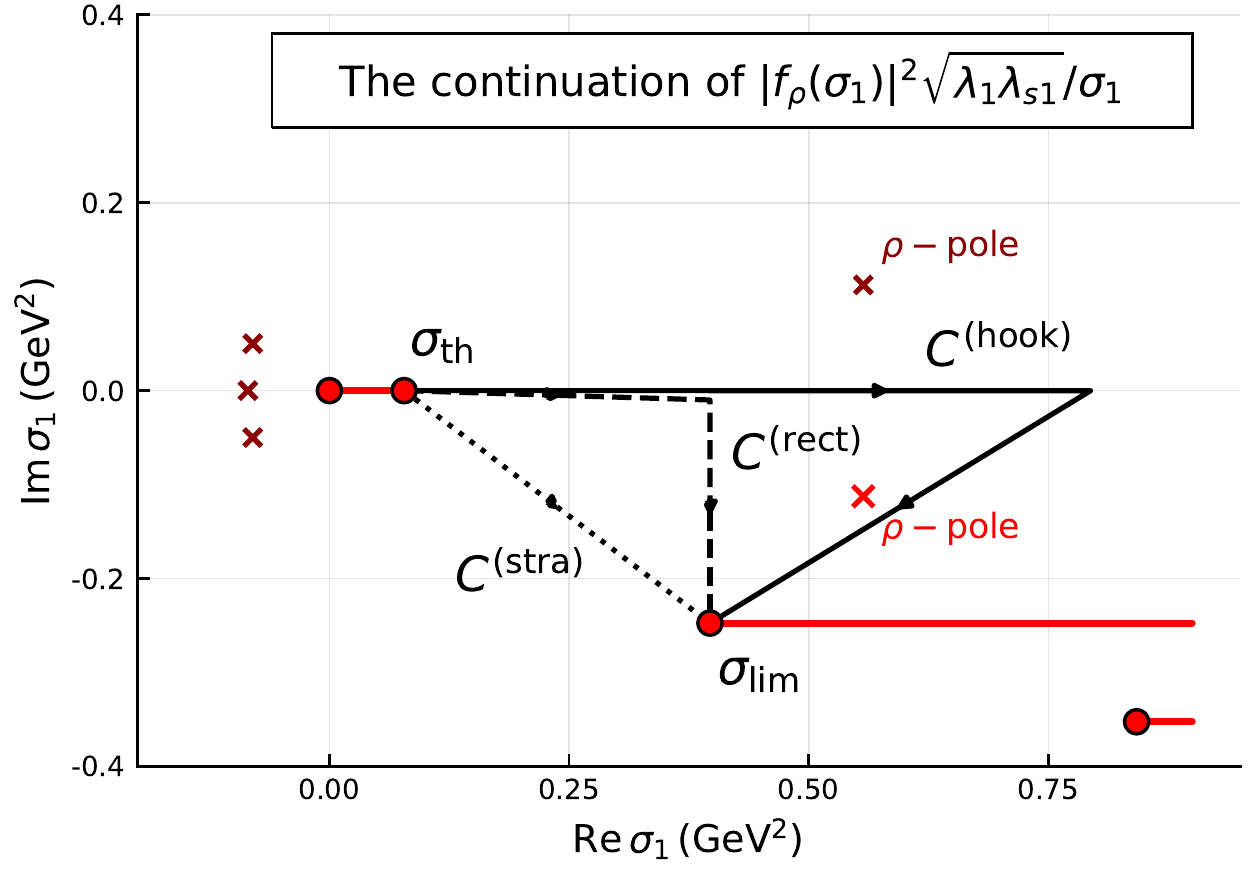}
  \includegraphics[width=0.45\textwidth]{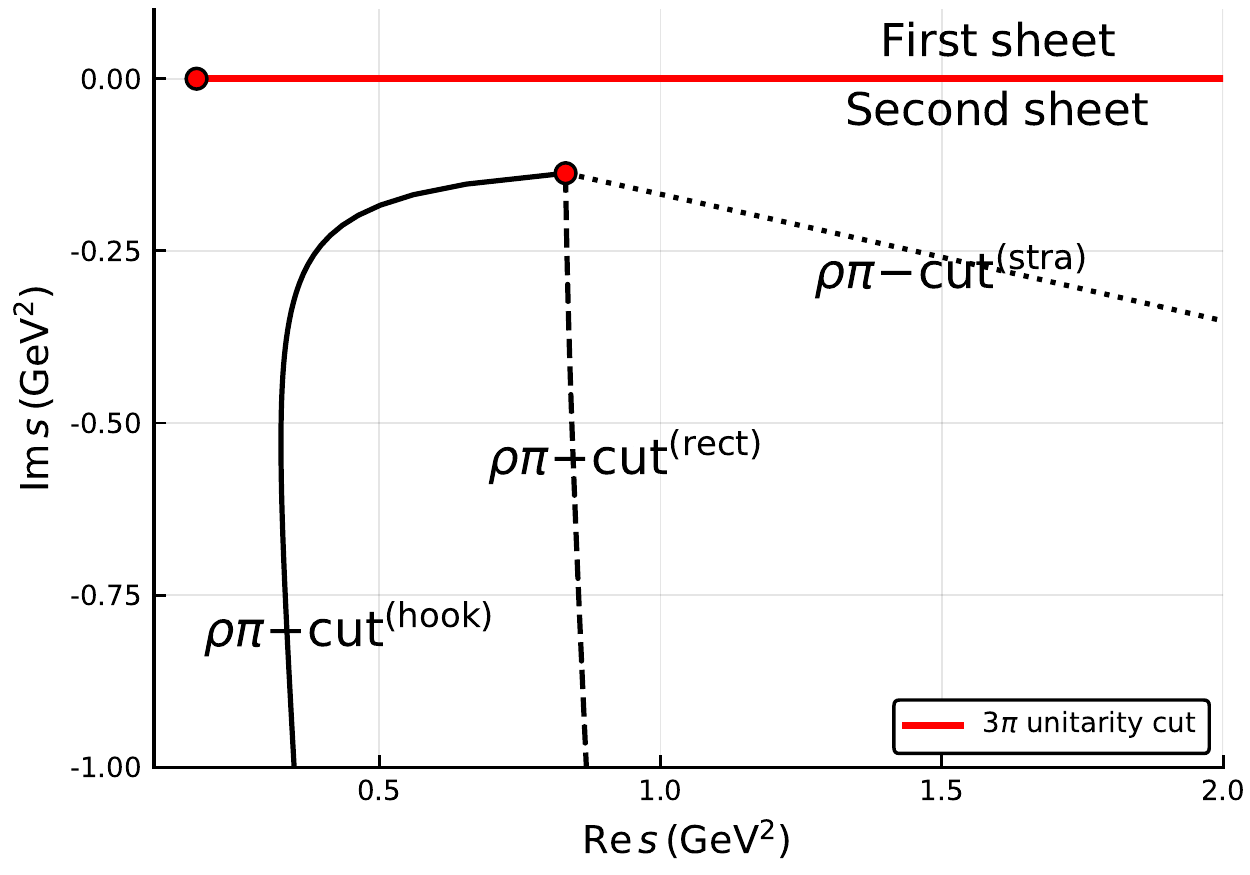}
  \caption{The left plot shows the complex plane of the integrand of Eq.~\eqref{eq:rho.qtb.complex}, for $s = 0.6-0.35i\,\GeV^2$.
  The red circular markers are the square-root branch points, the crosses indicate positions of the poles.
  The integration paths from Eq.~\eqref{eq:integration.paths.sigma1} are shown by the solid lines with arrows.
  The right plot presents the location of the $\rho\pi$ cut for the different integration paths.
  }
  \label{fig:qtb.sigma.integration}
\end{figure*}
For complex values of $s$, the integral for the $\rhoQTB(s)$ in Eq.~\eqref{eq:rho.qtb.complex} has the upper endpoint in the complex plane,
which requires a prescription for the path of integration.
The value of the integral does not depend on the path of integration, unless there are singularities of the integrand in the complex plane.
The integrand is plotted for complex values of $\sigma_1$ in Fig.~\ref{fig:qtb.sigma.integration}.
It has four branch points in the $\sigma_1$-variable: $0$, $\sTH = 4m_\pi^2$, $\sLIM = (\sqrt{s}- m_\pi)^2$,
and $(\sqrt{s}+ m_\pi)^2$, coming from the product of the K\"all\'en functions,\footnote{
The branch points are connected by cuts. Since the integral is calculated numerically it is
important to make sure that the integration path does not cross any cut between the integration end points.
To illustrate the cut choice shown in Fig.~\ref{fig:qtb.sigma.integration}, we write
\begin{equation*}
  \lambda^{1/2}(\sigma,m_\pi^2,m_\pi^2)\lambda^{1/2}(s,\sigma,m_\pi^2) = \sqrt{\sigma}\sqrt{\sigma-4m_\pi^2}\sqrt{(\sqrt{s}-m_\pi)^2-\sigma}\sqrt{(\sqrt{s}+m_\pi)^2-\sigma}.
\end{equation*}
For real values of $s$, this expression has two short branch cuts on the real axis: one between $0$ and $\sTH$, and the other between the points $(\sqrt{s}\pm m_\pi)^2$.
When $s$ is complex the first $s$-independent cut remains,
while the second one splits into two straight cuts to the right with the branching points $(\sqrt{s}\pm m_\pi)^2$ as shown in Fig.~\ref{fig:qtb.sigma.integration}.}
and the resonance pole of the $\rho$-meson at $\sigma_p = (m_\rho^\text{(pole)}-i\Gamma_\rho^\text{(pole)}/2)^2$.\footnote{
For the $\rho$-meson the pole parameters are very close to the Breit-Wigner parameters
$m_\rho^\text{(pole)} \approx m_\rho$, $\Gamma_\rho^\text{(pole)} \approx \Gamma_\rho$.}
Singularities of the integral arise when the upper integration endpoint touches one of the singularities of the integrand.
The $\rho$-meson pole in the integrand transforms into a branch singularity in the integral function.
We find the branch point $s_{\rho\pi}$ by checking when the upper integration endpoint touches
the $\rho$-meson pole; $s_{\rho\pi} = (m_\rho^\text{(pole)}+m_\pi-i\Gamma_\rho^\text{(pole)}/2)^2$.
It is indeed a branch singularity, because for every $s$ there are several ways to connect the integration limits in Eq.~\eqref{eq:rho.qtb.complex}
(see for example the solid and the dotted paths in the left panel of Fig.~\ref{fig:qtb.sigma.integration}) which yield integral values
differing by the residual of integrand in the $\rho$-meson pole.
Practically, the choice of the integration path determines the location of the $\rho\pi$ branch cut
in the complex $s$-plane as the loci of $s$ values, for which the integration path goes through the pole.
To demonstrate the evolution of the cut in the $s$-plane we consider the three different paths given in Eq.~\eqref{eq:stra}:
\begin{subequations}\label{eq:integration.paths.sigma1}
\begin{align}\label{eq:stra}
  C_\sigma^{(\text{stra})}: &\quad \sTH \to \sPTH\\ \label{eq:rect}
  C_\sigma^{(\text{rect})}: &\quad \sTH \to \re \sPTH \to \sPTH,\\\label{eq:hook}
  C_\sigma^{(\text{hook})}: &\quad \sTH \to 5\re \sPTH \to \sPTH.
\end{align}
\end{subequations}
The corresponding $\rho\pi$ cut locations are shown in the right panel of Fig.~\ref{fig:qtb.sigma.integration}.
The path $C_\sigma^{(\text{hook})}$ rotates the $\rho\pi$ cut such that it opens up a larger area of the closest unphysical sheet and
is used in the following for finding poles and illustration purposes.

\begin{figure*}
  \includegraphics[width=0.45\textwidth]{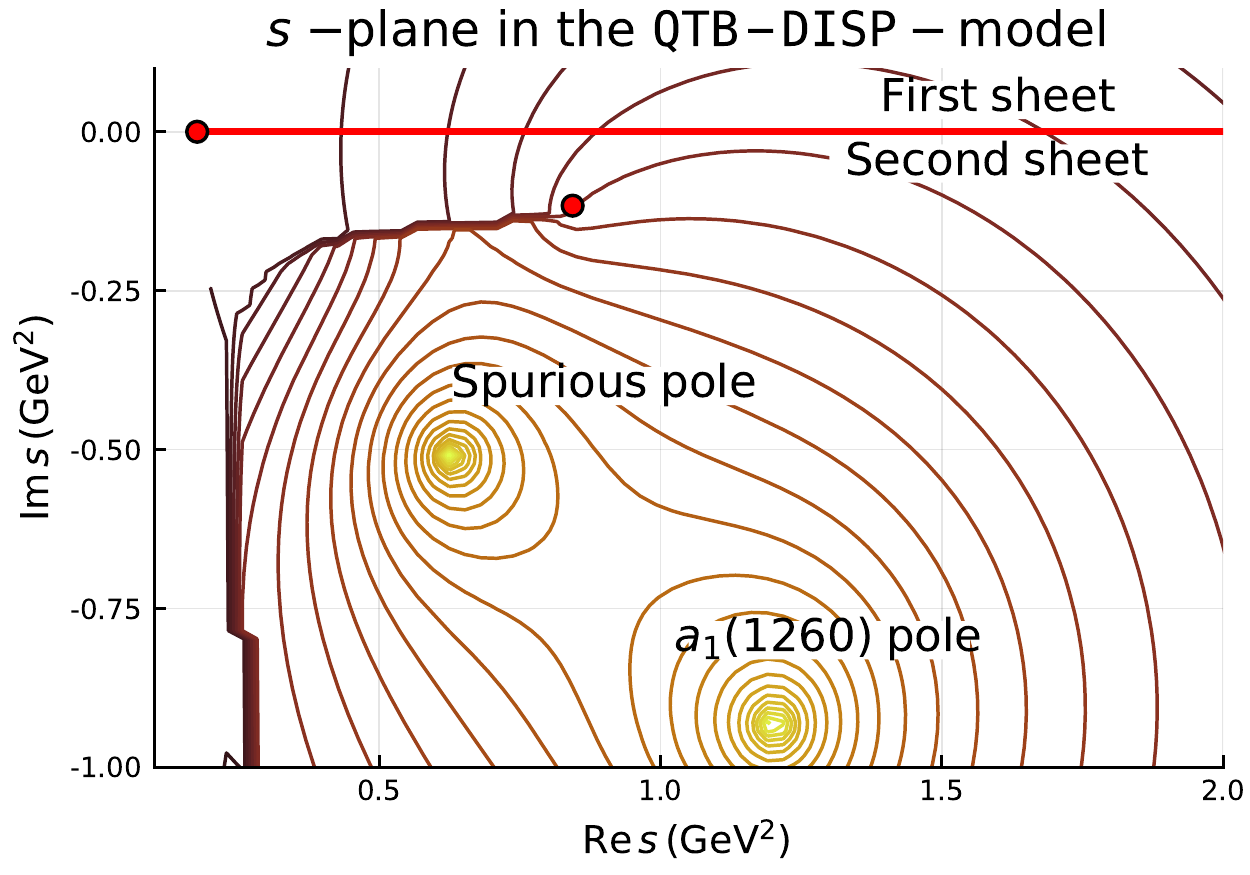}
  \includegraphics[width=0.45\textwidth]{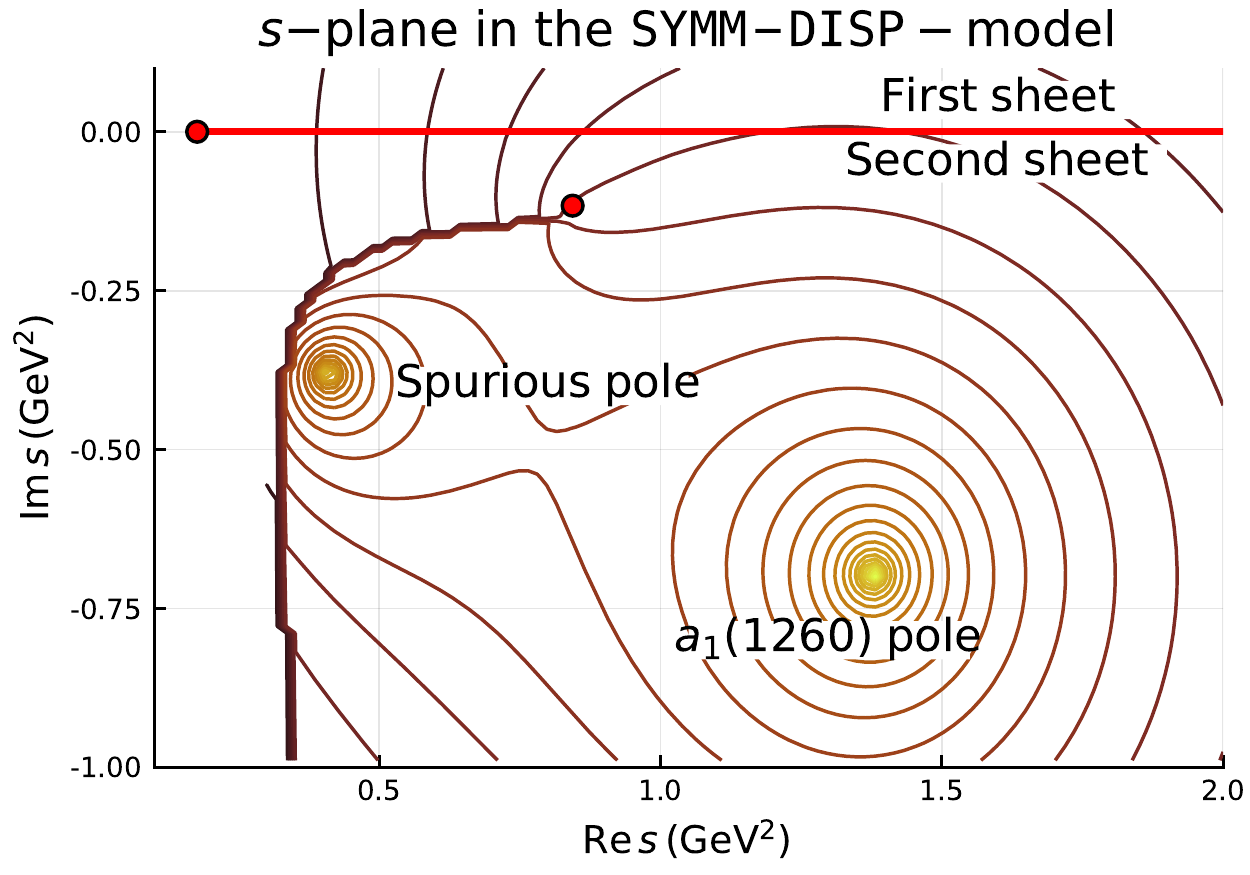}
  \caption{Analytic continuation of the amplitude $t(s)$ in Eq.~\eqref{eq:amp} for different models:
  \QTBDISP (Left plot), \SYMMDISP (Right plot).
  Lines indicate the $|t(s)|$ equipotential levels.
  The poles of the amplitude are the bright spots. The red dots indicate branch points corresponding to the opening of decay channels.
  }
  \label{fig:poles.continuation}
\end{figure*}

The amplitude $t(s)$ in the complex $s$-plane for the \QTBDISP model is shown in the left panel of Fig.~\ref{fig:poles.continuation}.
Naively, one would expect a single pole in the complex plane,
originating from the single $K$-matrix pole, $g^2/(m^2-s)$, present in Eq.~\eqref{eq:amp}.
In contrast to this expectation, two poles are observed. Furthermore, both are rather close to the physical region.
The correspondence between the $K$-matrix poles and the complex poles can be established by varying the coupling $g$.
In the limit $g\rightarrow 0$ the complex poles should approach the real axis at the position of the corresponding $K$-matrix poles.
We find that the deep pole approaches the real axis at $s = m^2 = (1223\,\MeV)^2$ (see Table~\ref{tab:models} with the fit results), while the left pole goes to $s = 0$.
Due to these observations, we identify the deep pole with $a_1(1260)$-pole label, \ie corresponding to a resonance,
and the left pole with a ``spurious''-pole, \ie, an artifact from our \parametrization in Eq.~\eqref{eq:amp}.
This exercise also helps us to understand the origin of the spurious pole: it is the $1/s$ singularity in $\rhoQTB$ (see Eq.~\eqref{eq:rho.qtb.complex}).
Clearly, this pole is an integral part of the model.
In Appendix~\ref{app:exploratory.studies} we consider variations of the model attempting to get rid of the spurious pole.
We show that its effect on the real axis is indeed required by the data. It effectively parameterizes
the unphysical sheet singularities, \eg the left-hand cuts related to the cross channel exchanges between pions in the final state.
For now, we conclude by extracting the positions of the $a_1(1260)$ resonance pole in the \QTBDISP model.
We use the convention $s_p=(m_p + i\Gamma_p/2)^2$, obtaining
\begin{equation} \label{eq:QTBDISP.result}
  \textrm\QTBDISP:\qquad m_p^{(a_1(1260))} = (1166 \pm 6)\,\MeV,\quad
  \Gamma_p^{(a_1(1260))} = (798 \pm 26)\,\MeV.
\end{equation}
For the error estimation we used the bootstrap technique~\cite{Press:1992zz, EfroTibs93}:
$1000$ sets of pseudo data were generated using
the original data and the covariance matrices, with the correlations  taken into account in the Gaussian approximation.
By refitting the pseudo datasets, we collect samples of the parameters, which we use to estimate their uncertainties.
The distributions of the mass and width of the pole obtained from the bootstrap are Gaussian to a good approximation.
The fit results and the calculated error ellipses are shown in Fig.~\ref{fig:a1.pole.systematic.studies}.
The mean values of the bootstrap sample for the pole positions differ from the real data fit results
by $<0.2\sigma$ which indicate a good consistency and negligible bias of the bootstrap method~\cite{EfroTibs93}.

\subsection{Analytic continuation of the \SYMMDISP model}
\label{sec:symm}
\begin{figure*}
  \includegraphics[width=0.45\textwidth]{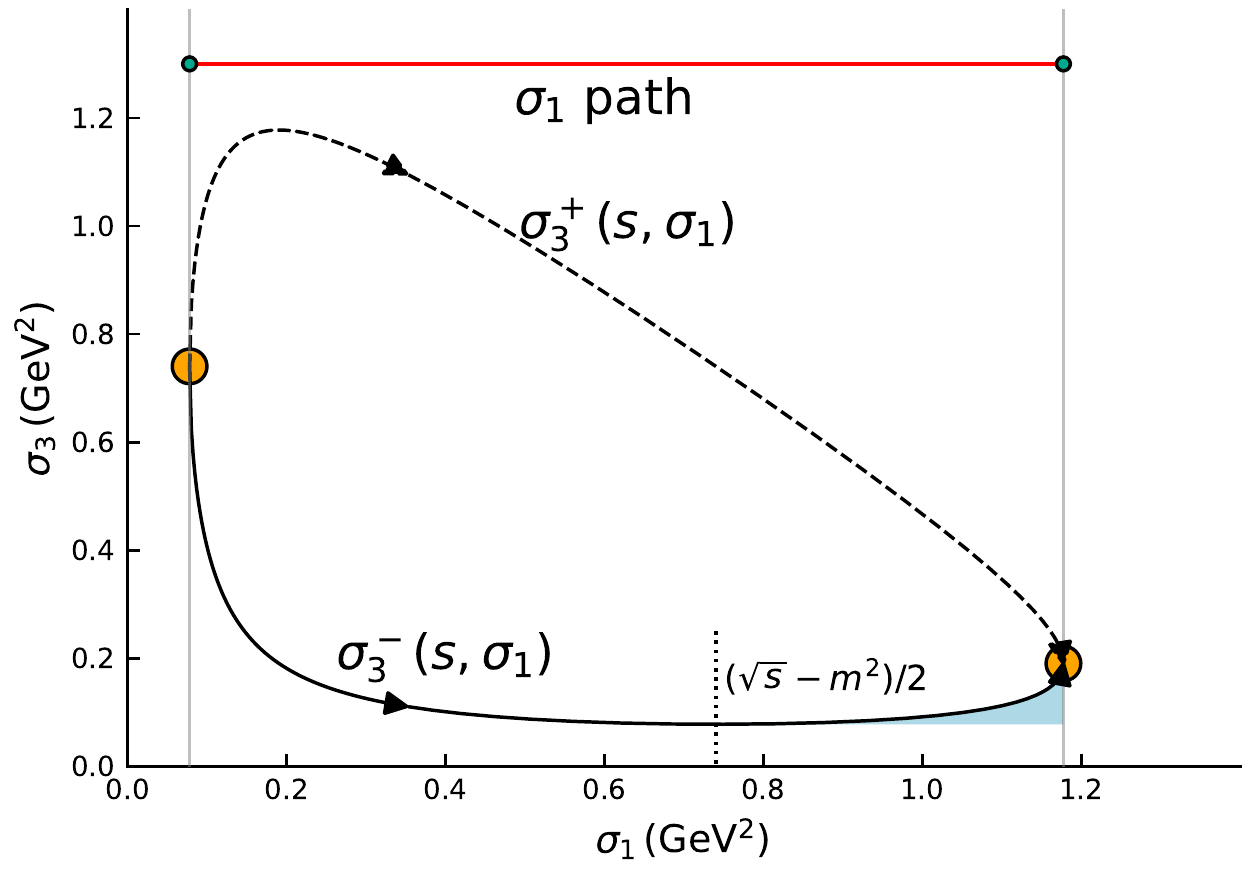}
  \includegraphics[width=0.45\textwidth]{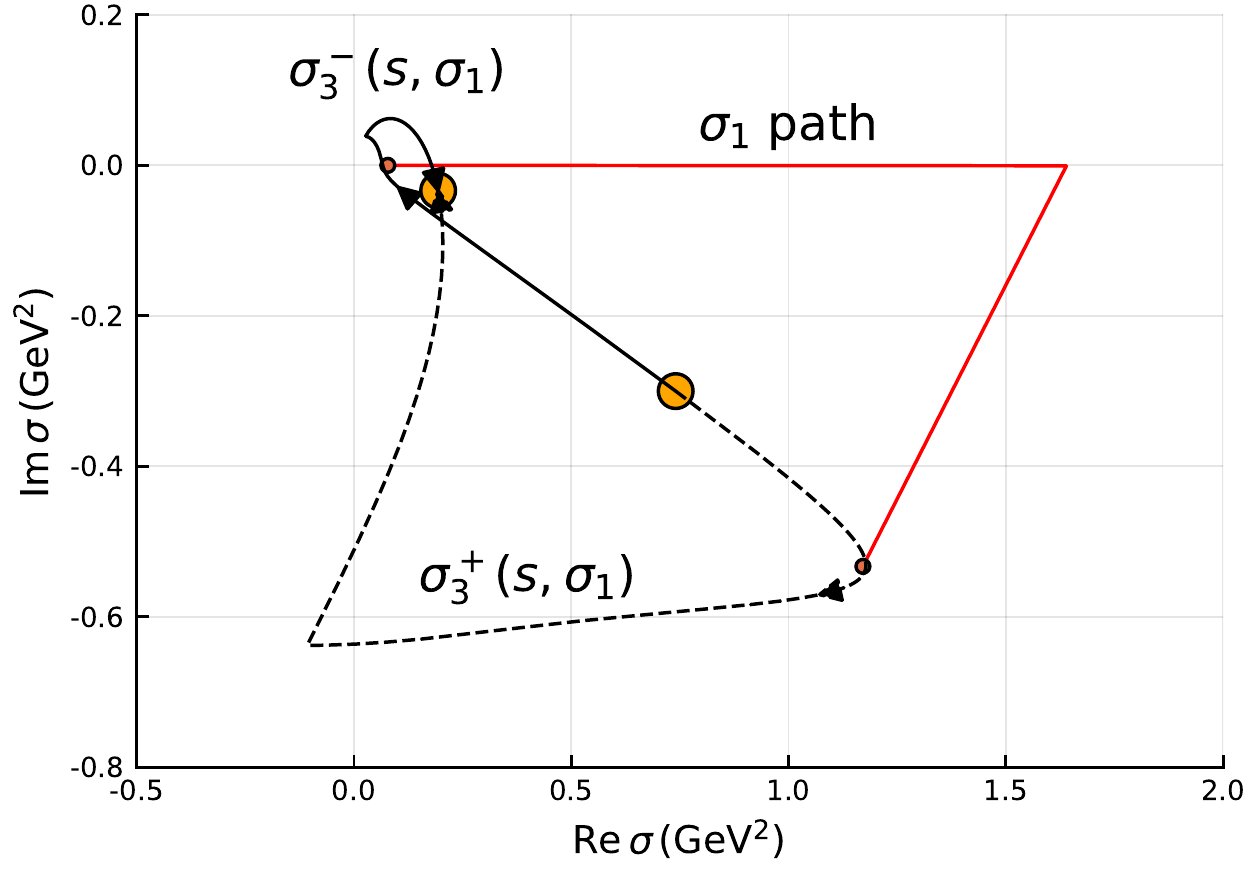}
  \caption{Integration paths in the complex $\sigma$-plane: while $\sigma_1$ is moving along the path $\sTH\to\sLIM$,
  the integration endpoints $\sigma_3^+(s,\sigma_1)$($\sigma_3^-(s,\sigma_1)$) are traveling in the complex plane along the lines shown by black solid (dashed) curve
  The left plot shows the integration ranges of $\sigma_1$ and $\sigma_3$ for a real value of $s = \unit{1.5}{\GeV^2}$.
  The red line is the straight integration path in $\sigma_1$.
  The yellow circles indicate the border of the integration domain when the integration endpoints in $\sigma_3$ coincide.
  In the right plot, the same lines are shown in the complex $\sigma$ plane combined for $\sigma_1$ and $\sigma_3$ when $s = \unit{1.5-0.6i}{\GeV^2}$.
  The points $4m_\pi^2$ and $(\sqrt{s}-m_\pi)^2$ are shown by the small orange dots.
  While $\sigma_1$ moves along the contour $C^{(\text{hook})}$ indicated by the red line,
  the integration limits $\sigma_3^{\pm}$ follow the dashed and the solid lines analogously to the left plot.
  The shaded area indicates the additional contribution to the phase-space integral discussed in Eq.~\eqref{eq:rho.int.cont}.
  }
  \label{fig:complex.paths}
\end{figure*}

The evaluation of the discontinuity given by Eq.~\eqref{eq:rho} for complex $s$ is more complicated
since the angular integrals cannot be solved completely, see Appendix~\ref{app:integrals}.
We start by casting $\rhoSYMM(s)$ in the form:
\begin{equation} 
\rhoSYMM(s) = \rhoQTB(s) - \rhoINT(s),
\end{equation}
where the first term in the sum is the phase-space factor in the \QTBDISP model,
the second term is the interference contribution given by Eq.~\eqref{eq:rhoInt.final}.
Substituting $\frho\to\frhoI$ and $\frhos\to\frhoII$ in Eq.~\eqref{eq:rhoInt.final} we get:
\begin{align} \label{eq:rho.int} 
    \rhoINT(s) &= \frac{1}{2\pi(8\pi)^2 s} \int_{4m_\pi^2}^{\sPTH} \diff \sigma_1
\int_{\sigma_3^-(\sigma_1,s)}^{\sigma_3^+(\sigma_1,s)} \diff \sigma_3
\frac{\frhoII(\sigma_1)}{\sqrt{\sigma_1-4m_\pi^2}}\,
\frac{\frhoI(\sigma_3)}{\sqrt{\sigma_3-4m_\pi^2}} \nonumber\\
&\qquad\times
\frac{\Wpoly}{((\sqrt{s}+\sqrt{\sigma_1})^2-m_\pi^2)((\sqrt{s}+\sqrt{\sigma_3})^2-m_\pi^2)}.
\end{align}
The function \WpolyOpt{a}{b}{c} is a multivariable polynomial defined in Eq.~\eqref{eq:Wpoly}.
Omitting constant factors, the function $\frho(\sigma)$ is given by
\begin{equation}\label{eq:rho.parametrization} 
\frho(\sigma) \propto \frac{\sqrt{F(\sigma)}}{m_\rho^2-\sigma-im_\rho\Gamma(\sigma)},
\quad
\Gamma(\sigma) \propto \frac{i\sqrt{4m_\pi^2-\sigma}}{\sqrt{\sigma}} F(\sigma),
\quad
F(\sigma) \propto \frac{\sigma-4m_\pi^2}{\sigma-4m_\pi^2 + 4/R^2}.
\end{equation}
A right-hand cut is introduced by $i\sqrt{4m_\pi^2-\sigma}$.
In addition, there are two branch points: one at $\sigma = 0$ from the phase space in the width $\Gamma(\sigma)$, and another one at
$\sigma = 4m_\pi^2-4/R^2$ due to the Blatt-Weisskopf factor in the numerator.
The break-up momentum singularity  $\sqrt{\sigma-4m_\pi^2}$ in the numerator of $f(\sigma)$ is canceled
by the same factor which arises from the angular function (see Eq.~\eqref{eq:rho.int}).
The \parametrization of $\frho(s)$ in Eq.~\eqref{eq:rho.parametrization}
contains $5$ poles, as one can count by the order of the polynomial which would give zeros of the denominator.
They correspond to the $\rho$-meson poles at $(m_\rho\pm i\Gamma_\rho/2)^2$,
and three spurious poles lying far away from the physical region as shown in Fig.~\ref{fig:qtb.sigma.integration}.
The integration endpoints of the $\sigma_3$ variable, $\sigma_3^{\pm}(\sigma_1,s)$, describe the border of the Dalitz plot for fixed value of $s$ (Fig.~\ref{fig:complex.paths}, left panel),
\begin{equation} \label{eq:sigma_3^pm} 
  \sigma^\pm_3(\sigma_1,s) = \frac{s+3m_\pi^2-\sigma_1}{2} \pm \frac{\sqrt{\lamO\lamsO}}{2\sigma_1}.
\end{equation}
As soon as $s$ becomes complex the endpoints depart from the real axis and move into the complex plane.
The trajectories of the $\sigma_3^\pm$ as functions of $\sigma_1$ moving from $4m_\pi^2$ to $(\sqrt{s}-m_\pi)^2$ are non-trivial.
As shown in Fig.~\ref{fig:complex.paths}, while $\sigma_1$ moves along the $C^{(\textrm{hook})}$ path (see Eq.~\eqref{eq:hook}),
the $\sigma_3^-$ circles around the branch point $4m_\pi^2$. When $\sigma_3$ crosses the unitarity cut, the sheet, on which it is evaluated, must be changed.
\begin{figure*}
  \includegraphics[width=0.45\textwidth]{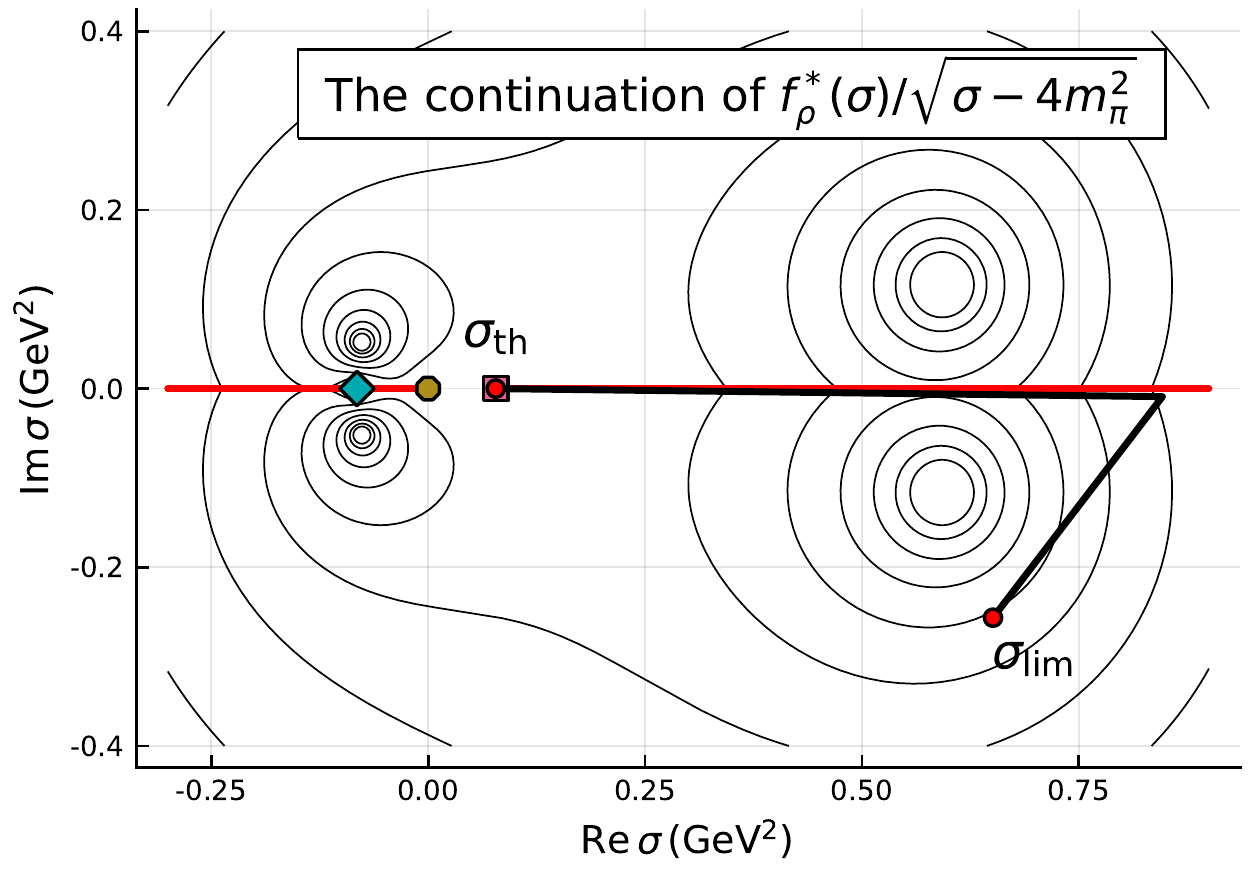}\,
  \includegraphics[width=0.45\textwidth]{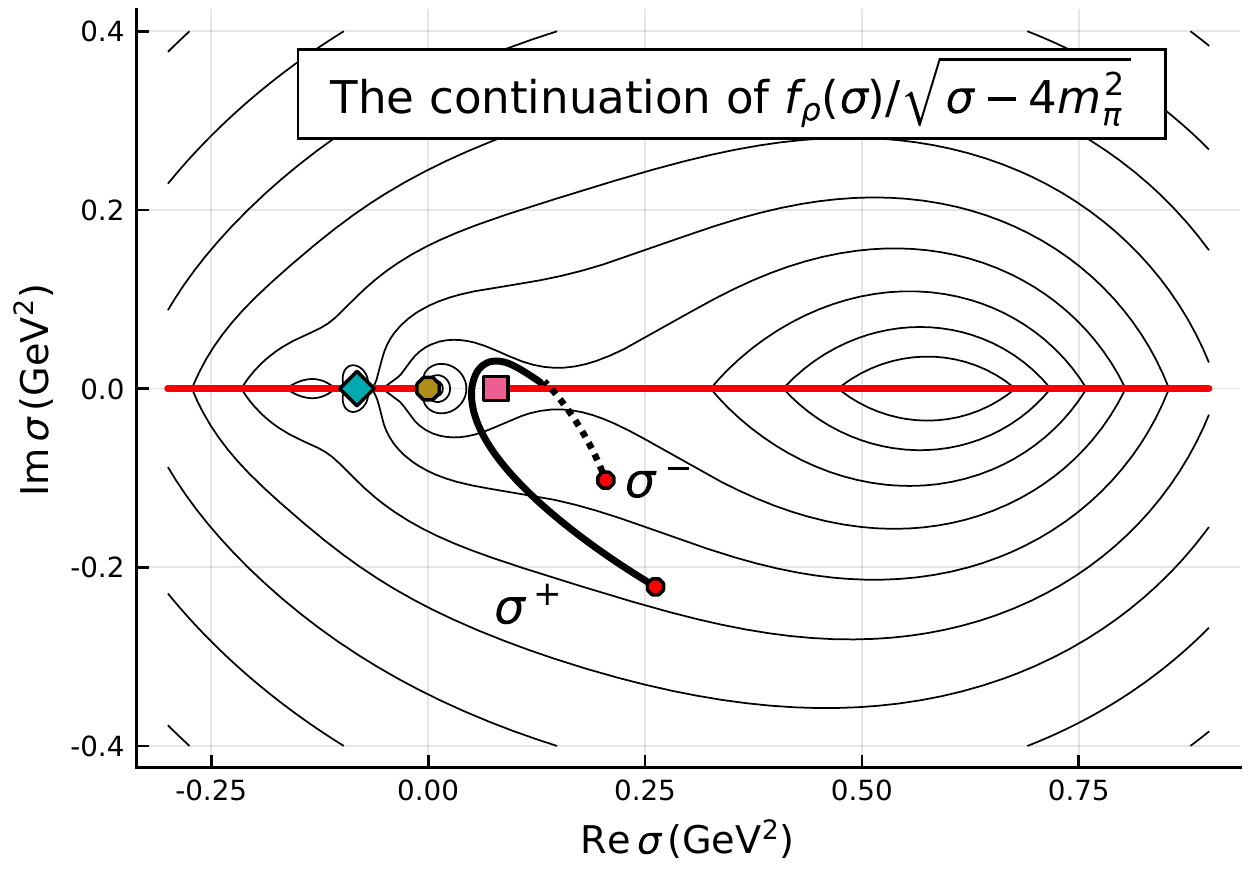}
  \caption{ The complex sheets of the isobar amplitude $\frho(\sigma)$ and $\frhos(\sigma)$.
  The left plot shows the analytic continuation of the function $\frho(\sigma)$ above and below the real axis.
  The function for positive imaginary part is the same as $\frhoI(\sigma)$; it is continuously connected to
  $\frhoII(\sigma)$ plotted for the negative imaginary part of $\sigma$.
  The right plot shows the analytic continuation of $\frhos(\sigma)$, where the sheets are inverted.
  The lines are $|\frho(\sigma)|$ equipotential surfaces. The circular spots are the poles
  (see also red crosses in the left plot of Fig.~\ref{fig:qtb.sigma.integration}).
  The markers on the real axis are the branch points of the left-hand cuts:
    the square marker shows the branch point from the break-up momentum located at $\sigma=4m_\pi^2$,
    the diamond marker the $\sigma=0$ branch point,
    the circular marker indicates the branch point related to the
      Blatt-Weisskopf factors in the numerator of the $\frho(\sigma)$ in Eq.~\eqref{eq:rho.parametrization}.
  }
  \label{fig:sigma.integration}
\end{figure*}
However, if the $\sigma_1$ path goes exactly through the point $(s-m_\pi^2)/2$, $\sigma_3^-$ just touches the branch point $4m_\pi^2$, (indeed, $\sigma_3^-((s-m_\pi^2)/2,s) = 4m_\pi^2$). In that case we are allowed to stay on the same sheet.
Therefore, there are two ways to calculate $\rhoINT$ for a complex argument (see Appendix~\ref{app:circling} for more details):
\begin{enumerate}
  \item $\rhoINT^{(1)}$: We choose a special path in $\sigma_1$,
  \begin{align}
    C_\sigma^{(\text{spec})}: &\quad \sTH \to (s-m_\pi^2)/2 \to \sPTH, \label{eq:spec}
  \end{align}
  the $\sigma_3^{\pm}$ always stay on the same sheet and can be connected with a straight (undistorted) path.
  \item $\rhoINT^{(2)}$: We let $\sigma_3^-$ circle around the branch point, changing sheets of $f(\sigma_3)$ appropriately.
  When $\sigma_1=\sTH$, the integration limits $\sigma_3^{\pm}$ coincide. For certain values of $\sigma_1$,
  $\sigma_3^-$ changes the sheet and, therefore, when $\sigma_1$ is in its upper limit $\sLIM$,
  the positions of $\sigma_3^{\pm}$ coincide, but they are on the different sheets.
  The integration path must be detoured around the branch point as shown in Fig.~\ref{fig:sigma.integration}.
\end{enumerate}
The first option provides a unique continuation of Eq.~\eqref{eq:rhoInt.final},
however, the integration contour is bound to pass through $(s-m_\pi^2)/2$ which is non-analytic point of the integrand (see Appendix~\ref{app:circling}).
The integrand in the second option stays analytic on the integration contour, but
in the limit of real $s$, the function $\rhoINT^{(2)}$ deviates from the original expression in Eq.~\eqref{eq:rhoInt.final}.
For $s = \re s + i\epsilon$, we change the sheet of $\sigma_3^-$ when $\sigma_1>(s-m_\pi^2)/2$, in contrast to the first option.
The mismatch is calculated by integrating the discontinuity across the $\sigma_3$ unitarity cut over the shaded area of Fig.~\ref{fig:complex.paths}.
\begin{align}\label{eq:rho.int.cont} 
  \rhoINT^{(1)}(s+i\epsilon)-\rhoINT^{(2)}(s+i\epsilon) &= \int_{(s-m_\pi^2)/2}^{(\sqrt{s}-m_\pi)^2} \diff \sigma_1
  \int^{\sTH}_{\sigma_3^-(\sigma_1,s)}
  \diff \sigma_3
  \frac{\left[\frhoI(\sigma_3+i\epsilon) - \frhoI(\sigma_3-i\epsilon)\right]}{\sqrt{\sigma_3-4m_\pi^2}}\\\nonumber
  &\qquad\times
  \frac{\frhoII(\sigma_1)}{\sqrt{\sigma_1-4m_\pi^2}}\,
  \frac{\Wpoly}{((\sqrt{s}+\sqrt{\sigma_1})^2-m_\pi^2)((\sqrt{s}+\sqrt{\sigma_3})^2-m_\pi^2)}.
\end{align}
The difference is practically negligible as shown in Fig.~\ref{fig:phase.space}.
The impact on the fit parameters and the values of the amplitude in the complex plane is a few orders of magnitude smaller than the statistical uncertainties. For the following discussion we use $\rhoINT^{(2)}(s)$ for the reason that the $\rho\pi$-cut can be rotated in the same way as before by using $C^{(\text{hook})}$ path in $\sigma_1$.
Interestingly, an analogous problem appears in relation to the Khuri-Treiman equations (see Appendix in Ref.~\cite{Aitchison:1966lpz}, Sec.~IV in Ref.~\cite{Pasquier:1968zz}). Ref.~\cite{Hwa.1963:xyz} gives arguments in favor of the first option.

As soon as the discontinuity is known for the whole complex plane, the amplitude on the unphysical sheet can be computed according to Eq.~\eqref{eq:tII-1=}.
The contour plot on the right panel of Fig.~\ref{fig:poles.continuation} presents the closest unphysical sheet of the amplitude, which is smoothly connected to the real axis.
We find two poles and identify them as the $a_1(1260)$ resonance pole and the left ``spurious'' pole as shown in Fig.~\ref{fig:poles.continuation}.
for the same reasoning as in Sec.~\ref{sec:qtb-cont}.
The pole parameters are
\begin{align}\label{eq:SYMMDISP.result} 
  \SYMMDISP: &\quad m_p^{(a_1(1260))} = (1209 \pm 4)\,\MeV,\quad
  \Gamma_p^{(a_1(1260))} = (576 \pm 11)\,\MeV.
\end{align}
The statistical errors are obtained from a bootstrap analysis as described above in Sec.~\ref{sec:qtb-cont}.
The combined results are presented in Fig.~\ref{fig:a1.pole.systematic.studies}.


\section{Systematic uncertainties}
\label{sec:systematic.studies}
\begin{figure*}
  \includegraphics[width=0.31\textwidth]{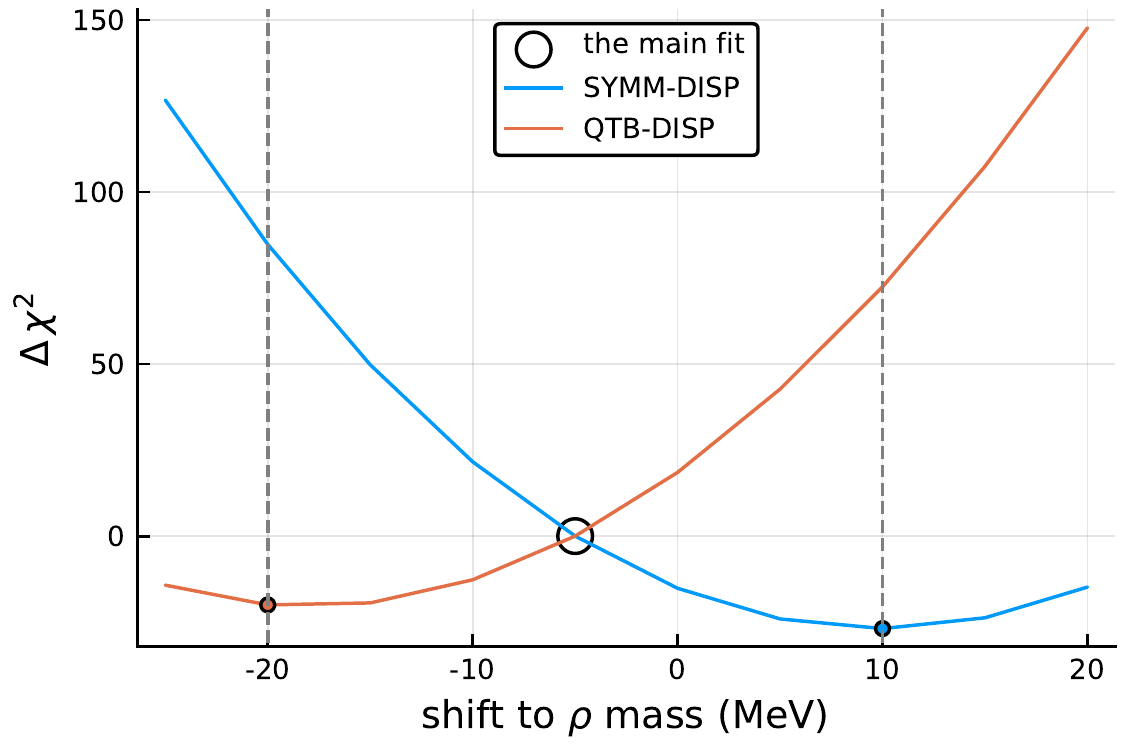}
  \includegraphics[width=0.31\textwidth]{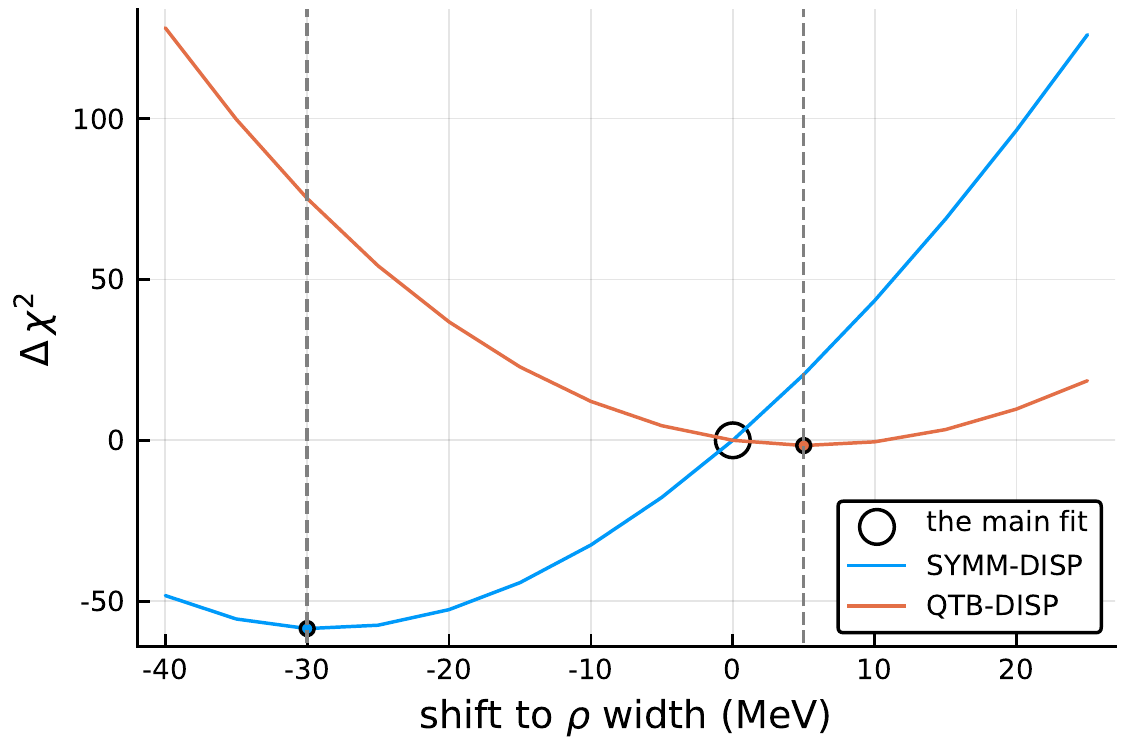}
  \includegraphics[width=0.31\textwidth]{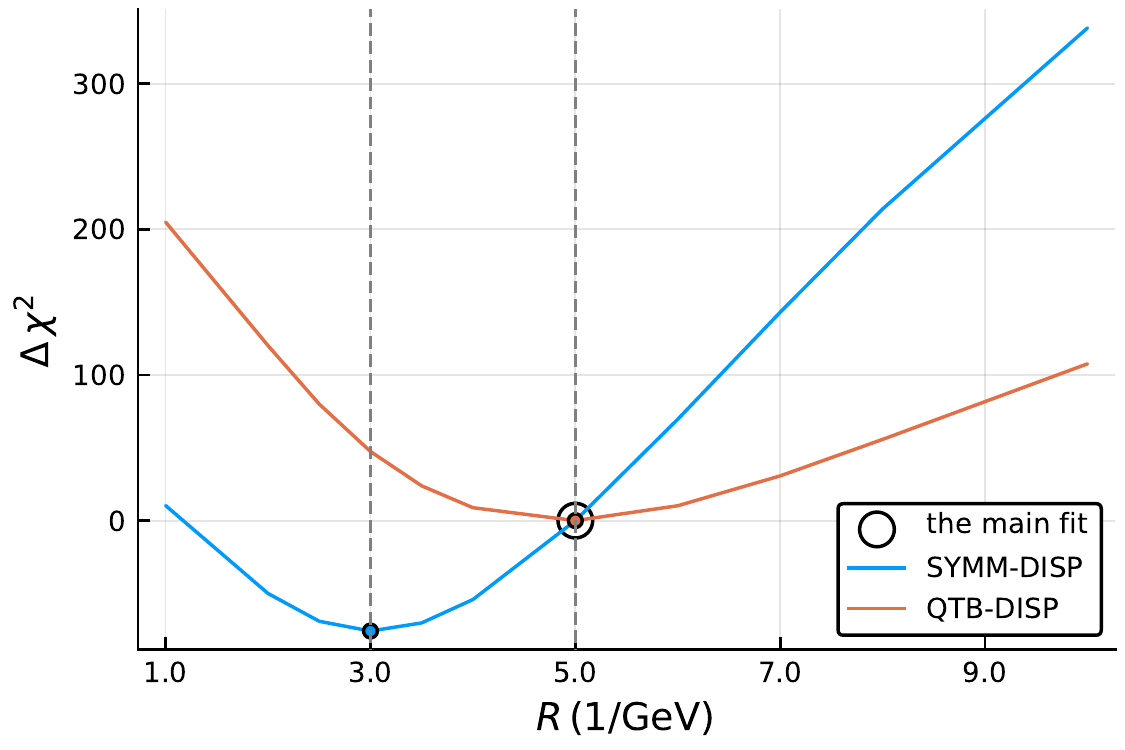}
  \caption{
  The change of the $\chi^2$ is plotted against the $\rho$-meson parameters in Eq.~\eqref{eq:isobar.parametrization}: the mass $m_\rho$,
  the width $\Gamma_\rho$ and the Blatt-Weisskopf size parameter $R$.
  The vertical lines indicate the estimated values where the minimum is found.}
  \label{fig:syst.scan}
\end{figure*}

\begin{figure*}[ht]
  \includegraphics[width=0.4\textwidth]{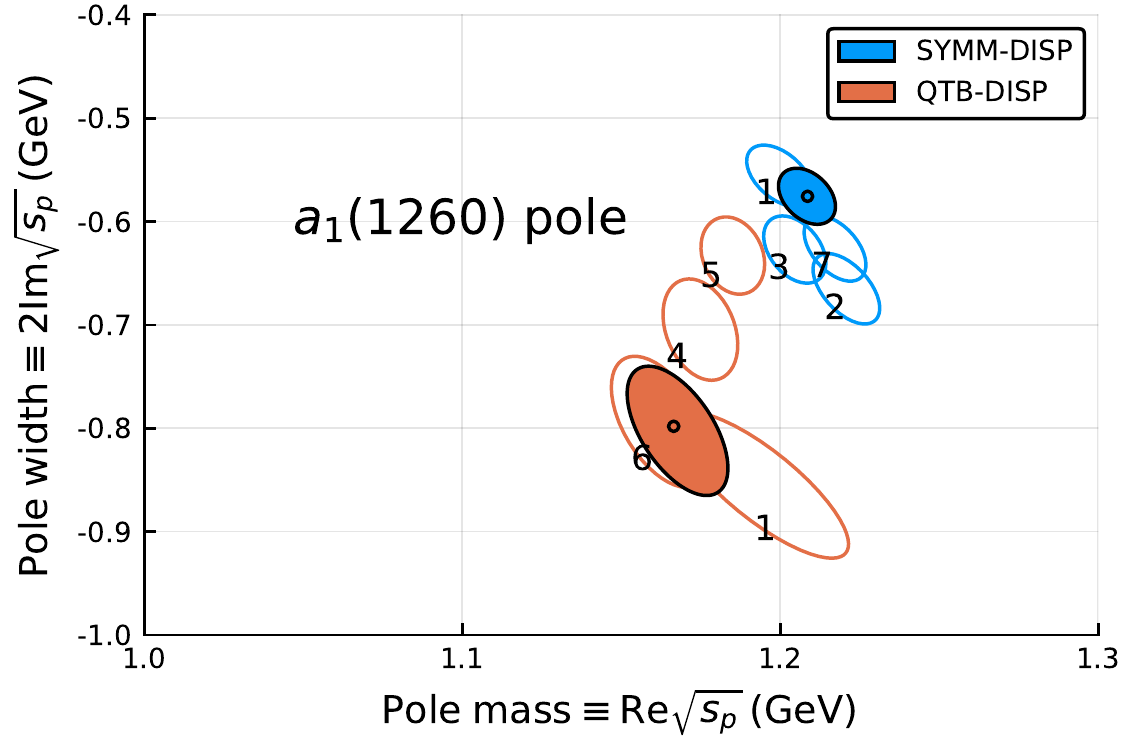}
  \caption{
  Extracted $a_1(1260)$ pole positions in the models \QTBDISP and \SYMMDISP.
  The ellipses show the $2\sigma$ contour of the systematic uncertainties obtained by the bootstrap method.
  The results of the systematic tests are shown by the open ellipses.
  The numerical labels correspond to the indices of the studies described in Table~\ref{tab:systematics.fit}.
  }
  \label{fig:a1.pole.systematic.studies}
\end{figure*}

The description of three-particle resonances is a difficult problem because of the complicated structure of final-state interactions,
which induces an interplay between different decay channels. The latter manifests itself in the modification of the isobar \lineshape
and the presence of interference terms.
The importance of three-body effects is readily seen in the difference of \SYMMDISP and \QTBDISP pole positions, \textit{cf.}\ \mbox{Eqs.~\eqref{eq:SYMMDISP.result}~and~\eqref{eq:QTBDISP.result}}.
Knowing that the interference between two $\rho\pi$ decay channels must be present, we now focus on systematic studies of \SYMMDISP , keeping \QTBDISP for a mere comparison.
The largest systematic uncertainty is the dependence of the $a_1(1260)$ pole position on the \lineshape of the subchannel resonance $\rho$.
In principle, we know that final-state interactions shift and skew the $\rho$ peak. 
The scale of the $\rho$-meson mass shift can be estimated from the studies of $\omega/\phi$ decays using Khuri-Treiman equations~\cite{Niecknig:2012sj,Danilkin:2014cra}.
Fig.~3 of Ref.~\cite{Niecknig:2012sj} suggests a shift of the real and imaginary parts of the isobar amplitude
of the order of $\unit{10}{\MeV}$ before and after final-state interactions are taken into account.
To estimate the effect on the $a_1(1260)$ pole position, we vary the parameters of $\frho(\sigma)$ in Eq.~\eqref{eq:isobar.parametrization},
\ie the mass $m_\rho$, the width $\Gamma_\rho$ and the  Blatt-Weisskopf radius $R$,
performing a $\chi^2$ scan over each parameter, while keeping the others
at their nominal values (Fig.~\ref{fig:syst.scan}).
The new pole position obtained for the parameter value which minimizes the $\chi^2$ for each scan is then used to estimate the systematic error for the pole position of the main fit.
The results of these studies are summarized in Table~\ref{tab:systematics.fit} (see fit studies \#$2$-$7$, were \#$4$ was introduced as an additional intermediate point outside of the minimum).
The $a_1(1260)$ pole position is extracted,
the results for the pole mass and width are represented in Fig.~\ref{fig:a1.pole.systematic.studies} by open ellipses.
\begin{table}
	\caption{The values $m$, $g$ and $\chi^2$ for fits described in Sec.~\ref{sec:systematic.studies}.
  For scans over parameters $m_\rho$, $R$ and $\Gamma_p$ we present the values of $m$, $g$ and $\chi^2$ obtained in the the minimum in the
  profile $\chi^2$ plots shown in Fig.~\ref{fig:syst.scan}.}
	\label{tab:systematics.fit}
    \begin{ruledtabular}
	\begin{tabular}{c| c || c|c|c || c|c|c }
    && & & \QTBDISP & & & \SYMMDISP \\\hline
    \#&Fit studies & $m$, \GeV & $g$, \GeV & $\chi^2/\textrm{n.d.f.}$ & $m$, \GeV & $g$, \GeV & $\chi^2/\textrm{n.d.f.}$ \\\hline
    1 & $s < \unit{2}{\GeV^2}$                            & $1.232$ &  $7.6$ & $53/62$   & $1.200$ & $6.57$ &  $ 81/62$ \\
    2 & $R' = \unit{3}{\GeV^{-1}}$                        &         &        &           & $1.211$ & $7.00$ &  $18/100$ \\
    3 & $m_\rho' = m_\rho + \unit{10}{\MeV}\,$            &         &        &           & $1.207$ & $6.85$ &  $83/100$ \\
    4 & $m_\rho' = m_\rho - \unit{10}{\MeV}\,$            & $1.204$ & $7.23$ &  $37/100$ &         &        &           \\
    5 & $m_\rho' = m_\rho - \unit{20}{\MeV}\,$            & $1.217$ & $7.01$ &  $30/100$ &         &        &           \\
    6 & $\Gamma_\rho' = \Gamma_\rho + \unit{5}{\MeV}$     & $1.223$ & $7.45$ &  $66/100$ &         &        &           \\
    7 & $\Gamma_\rho' = \Gamma_\rho - \unit{30}{\MeV}$    &         &        &           & $1.205$ & $6.79$ &  $36/100$
	\end{tabular}
    \end{ruledtabular}
\end{table}

We perform an additional test of the influence of heavier resonances, as the  $a_1(1640)$, by excluding the region $s>2\,$GeV$^2$ from the fit.
The fit quality does not change substantially, but get slightly worse due to the reduction of the degrees of freedom (see \#$1$ in Table~\ref{tab:systematics.fit}).
The values for the pole position are shown in Fig.~\ref{fig:a1.pole.systematic.studies} and included to the systematic error of our final result.

The final systematic uncertainties are found by assigning the maximal deviation of the pole position in the systematic studies to the main fit \SYMMDISP:
\begin{align*}
  \quad m_p^{(a_1(1260))} = (1209 \pm 4 ^{+12}_{-9})\,\MeV,&\quad
  \Gamma_p^{(a_1(1260))} = (576 \pm 11 ^{+89}_{-20})\,\MeV.
\end{align*}
where the first uncertainty is statistical and the second systematic.


\section{Conclusions}
\label{sec:conclusion}

In this paper we have presented a new analysis of the lightest iso-vector axial-vector resonance $a_1(1260)$ decaying to three charged pions.
Despite the fact that the corresponding $J^{PC}=1^{++}$ partial wave dominates the hadronic weak decay of $\tau$ leptons as well as diffractive reactions of high-energy pions, the parameters of the $a_1(1260)$ are still poorly known.
While the latter reactions suffer from an irreducible background due to non-resonant processes,
the system of three pions produced in $\tau$ decay provides a very clean access to axial-vector resonances.
Compared to a two-particle system, however, the system of three interacting particles exhibits additional phenomena,
such as 3-particle rescattering or interference between different decay chains.
These 3-body effects are taken into account using reaction models constraining the dynamics in the total invariant mass, however, without imposing subchannel unitarity.
We have considered four analytic models of an isolated resonance decaying to three pions via the $\rho\pi$ channel.
All these models satisfied \approximateUnitarity, but differ by the left-hand singularities and the treatment of the interference between
the two $\rho\pi$ decay channels.
Using the \taudecaycharged data from ALEPH~\cite{Schael:2005am},
we found that the dispersive models, having no left-hand singularities on the physical sheet, fit the data clearly better.

In order to find the pole position corresponding to the $a_1(1260)$ resonance, we have explored the analytic structure of the amplitude and
performed its analytic continuation into the complex plane of the \threepion invariant mass squared, a challenging, and technically demanding task,
requiring us to use a prescription for the integration paths in the two-pion invariant mass squared.
We have searched for the singularities in the closest unphysical sheet, and have identified a pole as the $a_1(1260)$ resonance.
The mass and width of the $a_1(1260)$ are given in terms of its pole position in the main \SYMMDISP model:
\begin{align*}
  \quad m_p^{(a_1(1260))} = (1209 \pm 4 ^{+12}_{-9})\,\MeV,&\quad
  \Gamma_p^{(a_1(1260))} = (576 \pm 11 ^{+89}_{-20})\,\MeV.
\end{align*}
The dominant source of systematic errors is the sensitivity to the details of the subchannel interactions.
The simplified \QTBDISP model, which neglects the interference between the two $\rho\pi$-channels, results in a significantly different pole position and a larger systematic uncertainty.

This analysis can be extended by further advancing the theoretical framework and constraining the model by fitting the Dalitz decay variables.
This will be possible when the data from BelleII or BESIII become available.
In addition, the results from this analysis will help to better constrain the non-resonant background in diffractive reactions, as measured by the COMPASS experiment.

\begin{acknowledgments}
This work was supported by the German Bundesministerium f\"{u}r Bildung und Forschung, 
the U.S.~Department of Energy under grants
No.~DE-AC05-06OR23177 
and No.~DE-FG02-87ER40365, 
Research Foundation -- Flanders (FWO), 
PAPIIT-DGAPA (UNAM, Mexico) under grants
No.~IA101717 and No.~IA101819, 
CONACYT (Mexico) under grant No.~251817, 
U.S.~National Science Foundation under award number
PHY-1415459, 
 Ministerio de Ciencia, Innovaci\'on y Universidades (Spain)
grants No.~FPA2016-77313-P and No.~FPA2016-75654-C2-2-P, and Universidad Complutense de Madrid.

\end{acknowledgments}

\newpage
\appendix

\section{Studies of the spurious pole}
\label{app:exploratory.studies}
\begin{figure*}
  \includegraphics[width=0.45\textwidth]{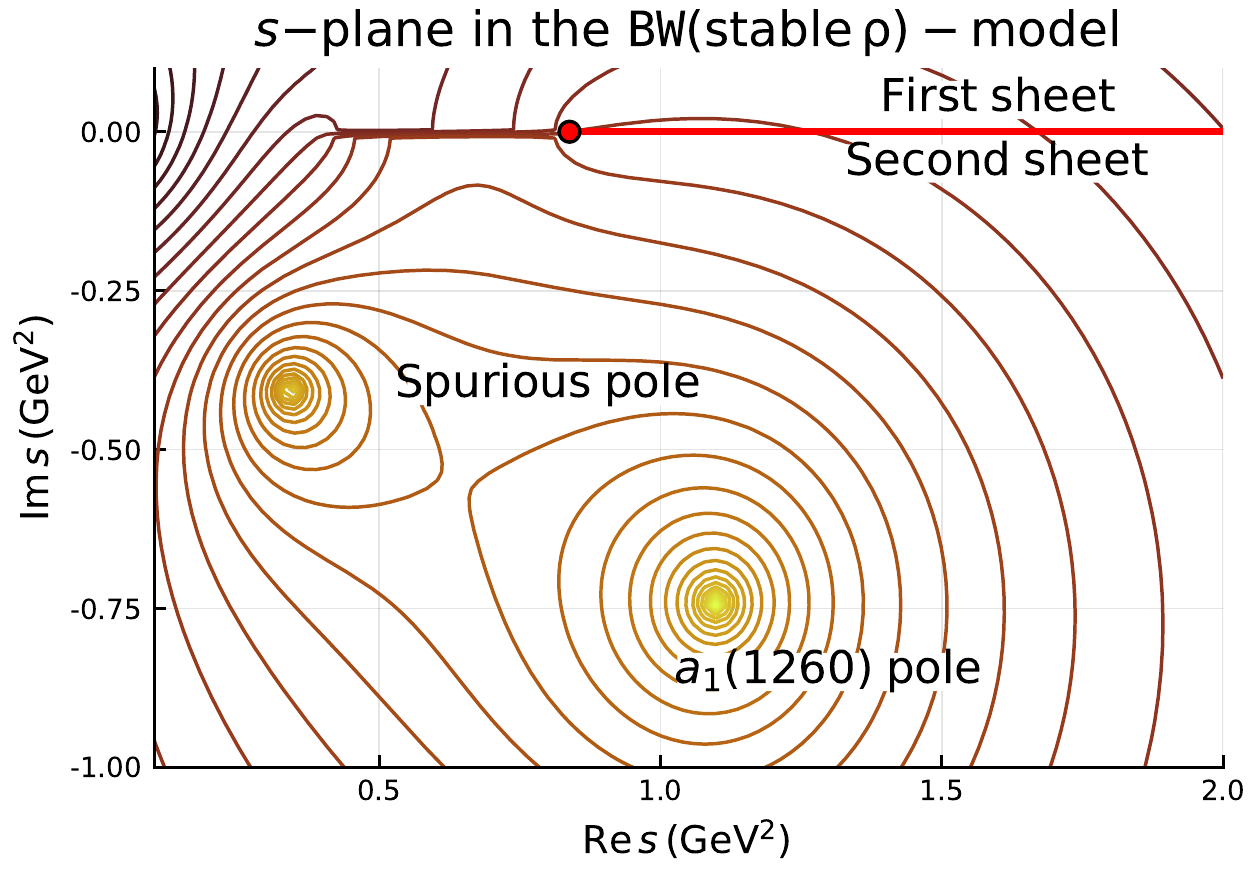}
  \caption{Analytic continuation of the amplitude $1/D_\text{BW}(s)$ from Eq.~\eqref{eq:app.DBW}.
  Lines indicate the $|D_\text{BW}|$ equipotential levels.
  The poles of the amplitude are the bright spots.
  The red dots indicate branch points for channel openings.
  }
  \label{fig:bw.continuation}
\end{figure*}

Performing the analytical continuation in Sec.~\ref{sec:poles} we have shown that, in addition to the expected $a_1(1260)$ pole,
there is a spurious pole rather close to the physical region.
At first, the spurious pole looks surprising, however, it is clearly present in every Breit-Wigner-like model of a resonance decaying to particles of different masses.
Indeed, the denominator of the Breit-Wigner amplitude with energy-dependent width  decaying to two scalar particles in an $S$-wave reads:
\begin{equation}\label{eq:app.DBW}
  D_\text{BW}(s) = m^2 - s - i m\Gamma(s), \quad \Gamma(s) = \frac{g^2}{16\pi m} \frac{\sqrt{(s-(m_1+m_2)^2)(s-(m_1-m_2)^2})}{s}.
\end{equation}
When $m_1 \neq m_2$, the equation $D_\text{BW}(s) = 0$ has $4$ complex roots,
which we can identify by the order of the polynomial which gives those roots:
\begin{equation}
  \left(16\pi s(m^2 - s)\right)^2 + g^4 \left(s-(m_1+m_2)^2\right)\left(s-(m_1-m_2)^2\right) = 0
\end{equation}
Since all coefficients of the polynomial are real, the poles appear in conjugated pairs above and below the real axis.
The two Breit-Wigner poles below the real axis are analogous to the $a_1(1260)$ and the spurious pole. 
To demonstrate this further, we draw the complex plane of the $1/D_\text{BW}(s)$ function with $m = 1.2\,$\GeV, $g = 7.8\,$\GeV, $m_1 = m_\rho$, $m_2 = m_\pi$
in Fig.~\ref{fig:bw.continuation}.
We find that the spurious pole has no influence on the physical region as long as the resonance is far from threshold and rather narrow.
Both poles become important for the real axis physics when the studied resonance is close to threshold or/and wider.

The spurious pole is a feature of Breit-Wigner-like models.
It is generated by the $1/s$ singularity of the phase space in Eq.~\eqref{eq:app.DBW}, and Eq.~\eqref{eq:rho.qtb.complex}.
In order to remove it, we try to exclude the $1/s$ factor from the dispersive term.
Following the studies of \QTBDISP, we consider a new model for scattering and production amplitudes $\hat t(s) = t(s)/s$ and $\hat a(s) = a(s)/s$,
and modify the unitarity equations accordingly.
\begin{subequations} \label{eq:unitarity.at.hat}
  \begin{align}
    2\im \, \hat t(s) &= \hat t^*(s) \,(s\rhoQTB(s))\, \hat t(s),\\
  	2\im \, \hat a(s) &= \hat t^*(s) \,(s\rhoQTB(s))\, \hat a(s),
  \end{align}
\end{subequations}
where $s\rho(s)$ is free of the $1/s$ singularity.
The \parametrization which satisfies the unitarity constraints is
\begin{equation} \label{eq:sQTB-DISPk}
  \hat{a}^{\sQTBDISP{k}}(s) = \frac{c'}{K^{-1}_k(s) - i s\rhoQTBtilde(s)/2},
\end{equation}
where the index $k$ gives the number of parameters in the function $K^{-1}_k(s)$,
the models are labeled $\sQTBDISP{k}$.
The function $s\rho(s)$ has a $\sim s^1$ asymptotic behavior, therefore
the dispersive integral must be subtracted twice.
The integrand is thus the same as in Eq.~\eqref{eq:rho.tilde},
but the integral is multiplied by an extra factor of $s$ as in Eq.~\eqref{eq:sQTB-DISPk}.
To make the dispersive integral independent of the subtraction points we must
consider a polynomial of order $k\ge 2$.
We consider three forms of functions $K_k(s)$,
\begin{align}
K_2(s) &= g^2/(m^2-s),\\
K_3(s) &= g^2/(s(m^2-s)+h)\\
K_4(s) &= g^2/(m^2-s) + h'/(m^{\prime 2}-s)
\end{align}
The $K_2(s)$ and $K_4(s)$ are inspired by the $K$-matrix approach with one and two poles, respectively,
while $K_3(s)$ is a special two-pole model which exactly coincides with $\QTBDISP$ when $h = 0$.
\begin{table*}
  \caption{Extension of Table~\ref{tab:models} with the models from Appendix~\ref{app:exploratory.studies}.
  We added the last column to present additional parameters which enter in the models.}
  \label{tab:models.spurious}
  \begin{ruledtabular}
  \begin{tabular}{c || c | c || r || c |c | c }
    Model  & $\rho(s)$ in the numerator & $C(s)$ in the denominator &
            $\chi^2/\text{n.d.f.}$ & $m$, \MeV & $g$, \GeV & $h$, $m^{\prime 2}$ \GeV$^{2}$ \\\hline
    \sQTBDISP{2}      & $\rhoQTB(s)$  & $\rhoQTBtilde(s)$   &  $979/100$ & $1.915$ & $17.94$ &--\\
    \sQTBDISP{3}      & $\rhoQTB(s)$  & $\rhoQTBtilde(s)$   &   $67/100$ & $1.075$ & $9.27$  & $0.578$\\
    \sQTBDISP{4}      & $\rhoQTB(s)$  & $\rhoQTBtilde(s)$   &   $42/100$ & $1.229$ & $6.01$  & $-39.3$, $0.0$\\
  \end{tabular}
  \end{ruledtabular}
\end{table*}

\begin{figure*}
  \includegraphics[width=0.45\textwidth]{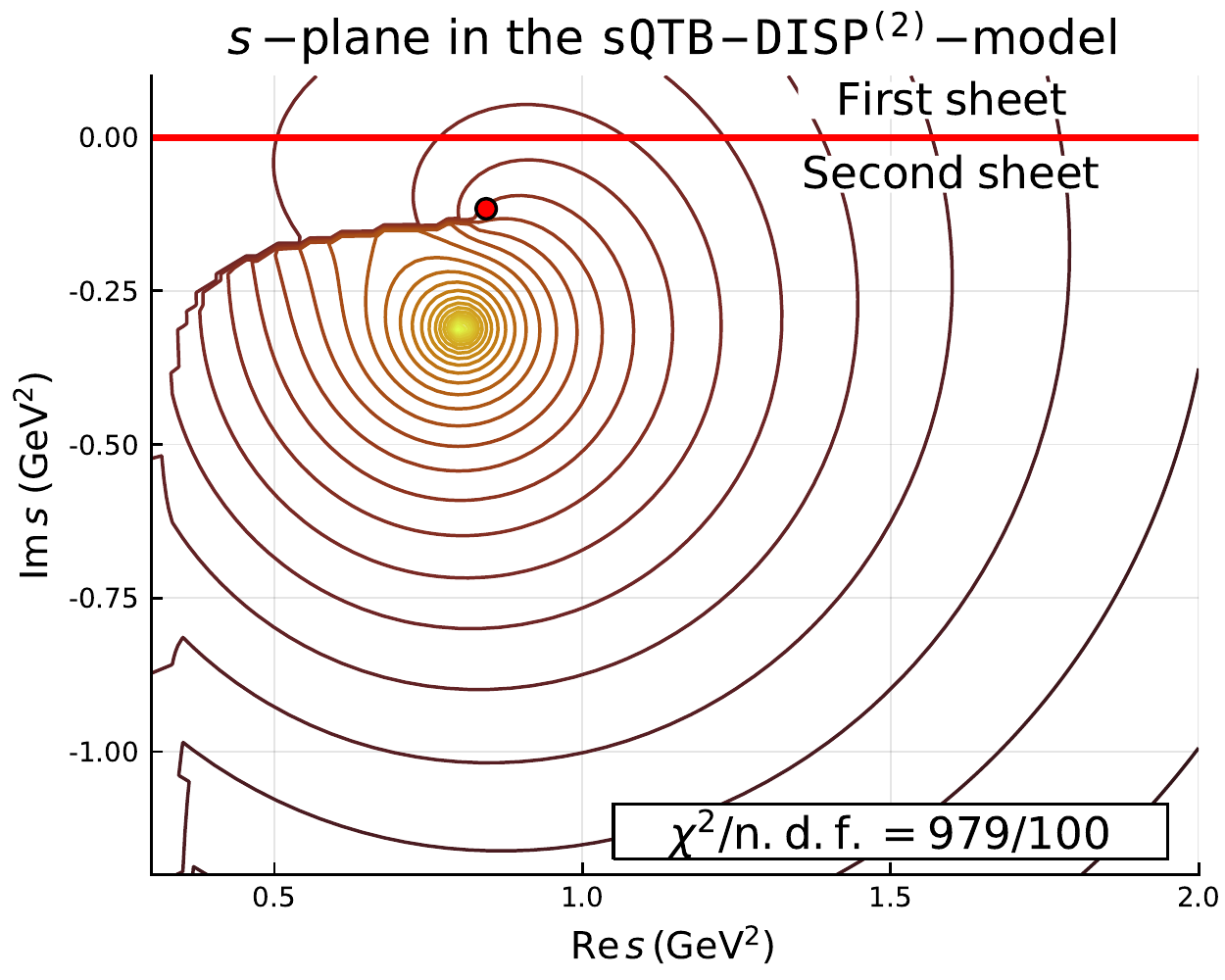}
  \includegraphics[width=0.45\textwidth]{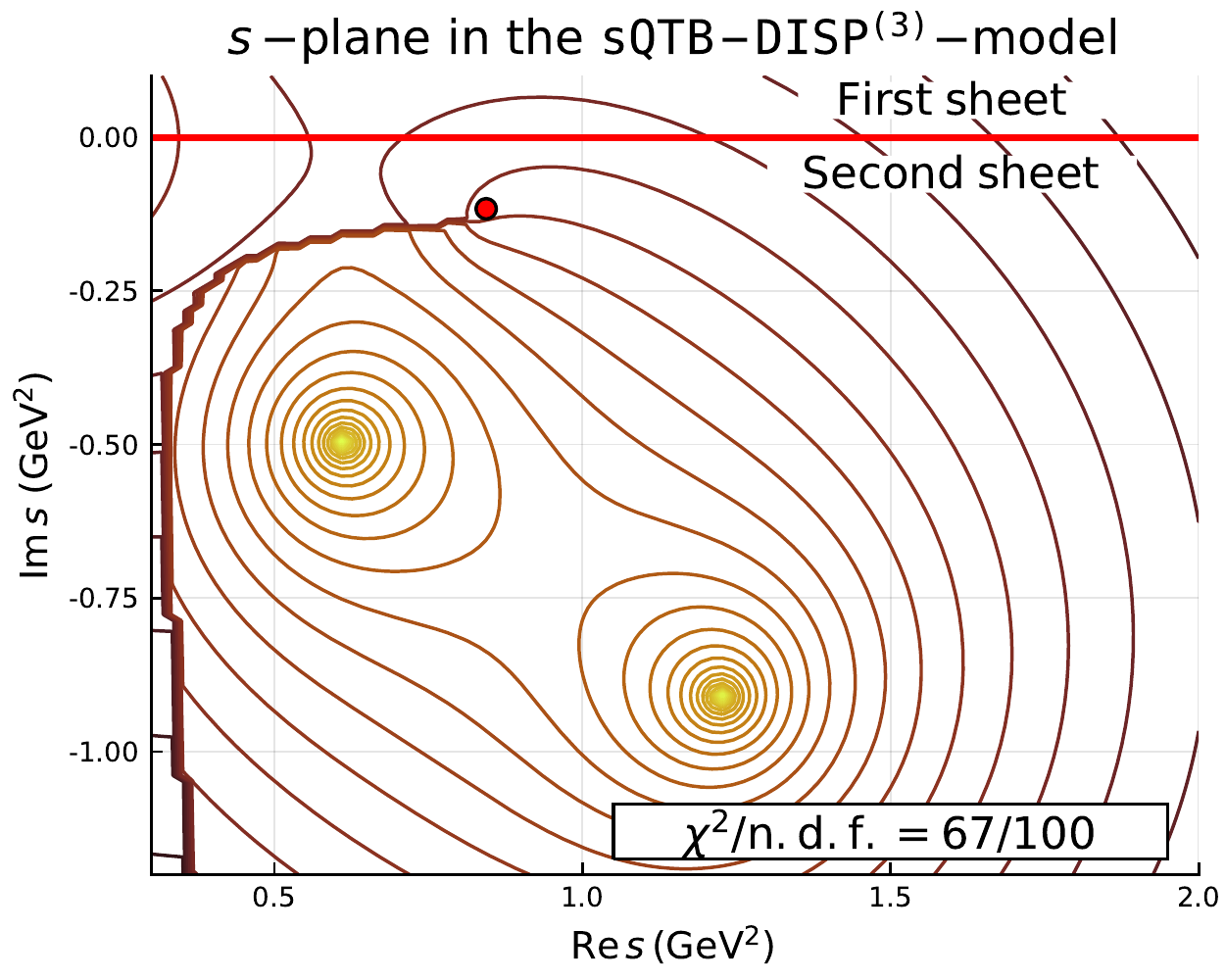}
  \caption{ $t(s)$ in the model \sQTBDISP{k}.
  Lines indicate equipotential levels for the $|\hat{t}^{\sQTBDISP{k}}(s)|$ function from Eq.~\eqref{eq:sQTB-DISPk}.
  The poles of the amplitude are the yellow spots. The red dots indicate branch points for channel openings:
  $3\pi$-branch point and $\rho\pi$-branch point.
  The complex plane for the model \sQTBDISP{2} (the models model \sQTBDISP{3}) fitted to the data is shown in the left (right) plot.
  The quality of the fit is indicated in the legend box on the right.
  }
  \label{fig:continuation.*s}
\end{figure*}

The models \sQTBDISP{k} are fitted to the data giving parameters presented in Table~\ref{tab:models.spurious}.
In Fig.~\ref{fig:continuation.*s} we show the continuation of the \sQTBDISP{2} model.
The spurious pole is no longer present.
However, the quality of the fit is not acceptable: the best $\chi^2/\text{n.d.f.}$ is equal to $979/100$.
When we increase the freedom by taking the model \sQTBDISP{3} the fit quality significantly improves to yield a $\chi^2/\text{n.d.f.} = 67/100$.
Quite spectacularly, the picture of the complex plane is changed back:
the place of the spurious pole is taken by the explicit pole introduced in the $K$-function (see the right plot of Fig.~\ref{fig:continuation.*s}).
The next relaxation of the setup in \sQTBDISP{4} overfits the data and gives $\chi^2/\text{n.d.f.} = 42/100$.
However, the positions of the resonance and spurious poles do not change much.

\begin{figure*}[ht]
  \includegraphics[width=0.95\textwidth]{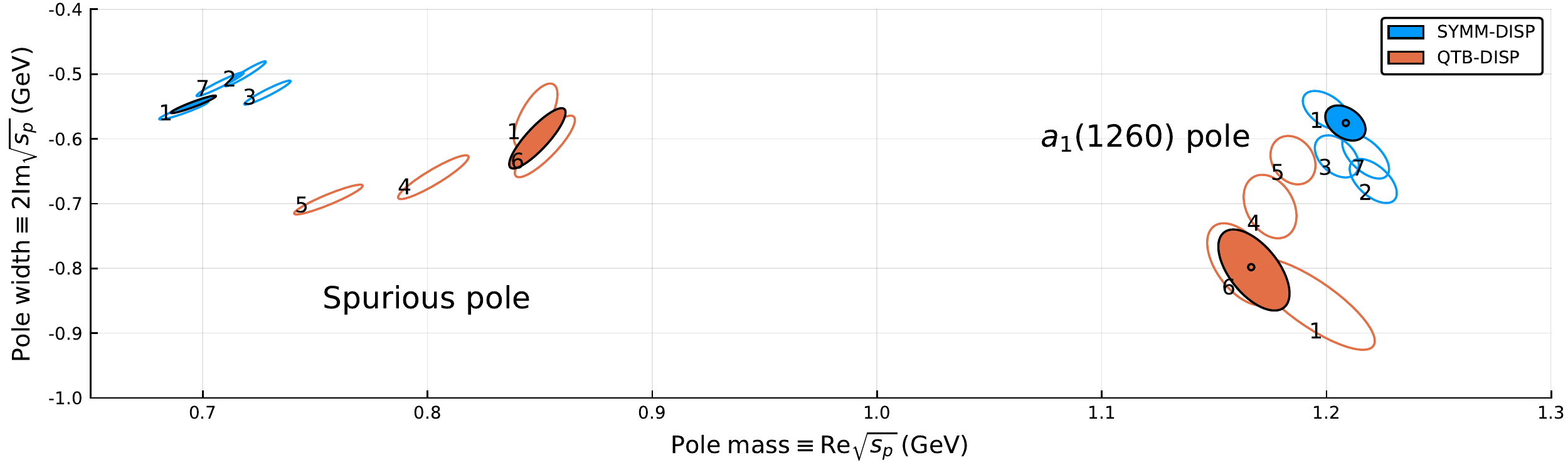}
  \caption{
  Extracted pole positions in the models \QTBDISP and \SYMMDISP: the resonance poles are on the right, the spurious poles are on the left.
  The ellipses show the $2\sigma$ contours of the statistical uncertainties obtained by the bootstrap method.
  The results of the systematic tests are shown by the open ellipses.
  The numerical labels correspond to the indexes of the systematic tests described in Table~\ref{tab:systematics.fit}.
  }
  \label{fig:poles.systematic.studies}
\end{figure*}

The position of the spurious pole was investigated for all systematic studies we performed in Sec.~\ref{sec:systematic.studies} as
shown in Fig.~\ref{fig:poles.systematic.studies}.


\section{Analytical simplification of the phase-space integral}
\label{app:integrals}
In this Appendix we demonstrate how the integrals in the phase-space factor $\rho(s)$ Eq.~\eqref{eq:rho} can be simplified using the properties of the Wigner $D$-functions.
\begin{equation}\label{eq:app.rho}
	\rho(s) = \frac{1}{2}\int\diff\Phi_3 \left|\frho(\sigma_1) N_0(\Omega_1,\Omega_{23}) - \frho(\sigma_3)N_0(\Omega_3,\Omega_{12})\right|^2,
\end{equation}

We start by explicitly defining the angles in the decay functions $N_0(\Omega_k, \Omega_{ij})$ given by Eq.~\eqref{eq:decay.function}.
The \threepion center-of-mass (CM) frame is oriented by the direction of $W$ in $\tau$ decay ($W$ helicity frame).
The momentum vector of the $\tau$ defines the $xz$ plane, \aka the production plane.
$\Omega_k = (\theta_k,\phi_k)$ denotes the polar and azimuthal angles of the vector $\vec{p}_i+\vec{p}_j$ in the CM-frame.
The $\Omega_{ij} = (\theta_{ij},\phi_{ij})$ are the spherical angles of the pion $i$ in the helicity frame of the isobar $(ij)$.
This helicity frame is obtained from the CM frame by active rotation $R^{-1}(\Omega_{k})$ and boost along the $z$-axis.
Equivalently, we can notice that the boost does not change azimuthal orientation, therefore,
the $y$-axis direction $\vec e_y$ in the helicity frame can be found by $\vec{e}_{z}^{\,\prime}\times \vec{e}_z$, where $\vec{e}_{z}^{\,\prime}$ is the original orientation of the CM $z$-axis.

We write the phase-space differential through the two pairs of spherical angles.
\begin{equation}
    \diff \Phi_3 = \frac{\diff \sigma_1}{2\pi}\,\frac{1}{8\pi}\frac{\sqrt{\lamsO}}{s}\frac{\diff \Omega_1}{4\pi}
    \,\,\frac{1}{8\pi}\frac{\sqrt{\lamO}}{\sigma_1}\frac{\diff\Omega_{23}}{4\pi}
     = \frac{\diff \sigma_3}{2\pi}\,\frac{1}{8\pi}\frac{\sqrt{\lamsT}}{s}\frac{\diff \Omega_3}{4\pi}
     \,\,\frac{1}{8\pi}\frac{\sqrt{\lamT}}{\sigma_3}\frac{\diff\Omega_{12}}{4\pi},
\end{equation}
where we used $\lambda_i = \lambda(\sigma_i,m_\pi^2,m_\pi^2)$, $\lambda_{si} = \lambda(s,\sigma_i,m_\pi^2)$, with $\lambda(x,y,z) = x^2 + y^2 + z^2 - 2(xy + yz + zx)$ the K\"all\'en triangle function.
The decay functions are conveniently normalized:
\begin{equation}
  \int \frac{\mathrm{d}\Omega_k}{4\pi} \frac{\mathrm{d}\Omega_{ij}}{4\pi}
  \left| N_0(\Omega_k,\Omega_{ij}) \right|^2 = 1.
\end{equation}
Now we can expand the squared expression in Eq.~\eqref{eq:app.rho}, use the normalization property, and combine the squared terms,
\begin{equation} \label{eq:split.rhosymm}
  \rhoSYMM(s) = \underbrace{\frac{1}{2\pi (8\pi)^2 s}\int \left|\frho(\sigma_1)\right|^2 \frac{\sqrt{\lamO\lamsO}}{\sigma_1}\,\diff \sigma_1}_{\rhoQTB(s)}
  -\underbrace{\int \diff \Phi_3 \frho(\sigma_1) \frhos(\sigma_3) \, N_0(\Omega_1,\Omega_{23})N_0^*(\Omega_3,\Omega_{12})}_{\rhoINT(s)},
\end{equation}
where we used the observation that the last integral is real. Indeed,
the term transforms to itself under complex conjugation due to the $1\leftrightarrow 3$ symmetry of the differential phase space
and the relation $N_0(\Omega_k, \Omega_{ij}) = -N_0(\Omega_k, \Omega_{ji})$ following from
\begin{equation}
D^l_{\lambda0}(\Omega_{ji}) = d^l_{\lambda0}(\pi-\theta_{ij})e^{-i\lambda(\pi+\phi_{ij})} =
(-1)^{l+\lambda} d^l_{\lambda0}(\theta_{ij})\,(-1)^{\lambda} e^{-i\lambda \phi_{ij}} = (-1)^{l}\,D^l_{\lambda0}(\Omega_{ij}),
\end{equation}
for $l = 1$, which has to be used for both terms $N_0(\Omega_3,\Omega_{12})$ and $N_0(\Omega_1,\Omega_{23})$.

The interference term can be further simplified by integrating over three angular variables.
The $N_{\Lambda}(\Omega_3,\Omega_{12})$ contains a product of Wigner $D$-functions which can be written as
\begin{equation}
  D^{1*}_{\Lambda\lambda}(\Omega_3)D^{1*}_{\lambda0}(\Omega_{12}) =
  D^{1*}_{\Lambda\lambda}(\phi_3,\theta_3,\phi_{12})\,d^{1}_{\lambda0}(\theta_{12}) =
  \sum_{\lambda'} D^{1*}_{\Lambda\lambda'}(\phi_1,\theta_1,\phi_{23}) d_{\lambda'\lambda}^1(\hattheta)\,
  d^1_{\lambda0}(\theta_{12}),
\end{equation}
where $\hattheta$ is the angle between $\vec p_1$ and $\vec p_3$ in CM-frame.
One can understand the relation in the following way.
The $D^{1*}_{\Lambda\lambda}(\Omega_3) = D^{1*}_{\Lambda\lambda}(\phi_3,\theta_3,0)$ and
$D^{1*}_{\lambda0}(\Omega_{12}) = D^{1*}_{\lambda0}(\phi_{12},\theta_{12},0)$ represent the rotations
$[R_z(\phi_{3})R_y(\theta_{3})]^{-1}$ and $[R_z(\phi_{12})R_y(\theta_{12})]^{-1}$.
The first transformation rotates the $3\pi$ system in the CM-frame such that the momentum $\vec p_1+\vec p_2 = -\vec p_3$ is aligned to the $z$-axis.
When the system is boosted to the $(12)$ rest frame (helicity frame), the second transformation aligns $\vec p_1$ to $z$-axis
(we remind that $\Omega_{12}$ stands for the spherical angles of the particle $1$ in the $(12)$ helicity frame).
Since the rotation $R_z(\phi_{12})$ commutes with the boost along $z$-axis, we can combine the three rotations in CM-frame,
$R_z^{-1}(\phi_{12})R_y^{-1}(\theta_{3})R_z^{-1}(\phi_{3})$.
The combined transformation has a clear meaning: it brings the $3\pi$ system to the $x-z$ plane such that $\vec p_3$ points to $-z$-direction.
The transformation $R_z^{-1}(\phi_{23})R_y^{-1}(\theta_{1})R_z^{-1}(\phi_{1})$ also brings the $3\pi$ system to the $xz$-plane while $\vec p_1$ is aligned with $-z$-direction.
The difference between the results of the transformations is a rotation about $y$-axis, represented by $d_{\lambda'\lambda}^1(\hattheta)$.
In that way, the function $D_{\Lambda \lambda}^{*1}(\phi_1,\theta_1,\theta_{23})$ appears in both decay functions in the interference term in Eq.~\eqref{eq:split.rhosymm}.
This allows us to solve three angular integrals analytically. The expression for $\rhoINT$ is simplified as follows:
\begin{equation}\label{eq:rhoInt.through.ddd}
  \rhoINT(s) = \frac{1}{(8\pi)^2 s} \int \frac{\diff \sigma_1}{2\pi} \frac{\diff\cos\theta_{23}}{2}
  \frhos(\sigma_3) \frho(\sigma_1)
  \frac{\sqrt{\lamO\lamsO}}{\sigma_1}\,
  \sum_{\lambda,\lambda'}
  d^1_{\lambda0}(\theta_{23}) d^1_{\lambda\lambda'}(\hattheta) d^1_{\lambda'0}(\theta_{12}).
\end{equation}
All angles can be expressed through the invariants,
\begin{align}
  \cos\theta_{23} &= \frac{\sigma_1(\sigma_3-\sigma_2)}{\sqrt{\lamO\lamsO}},&
  \sin\theta_{23} &= \frac{2\sqrt{\sigma_1}\sqrt{\phi(s,\sigma_1,\sigma_3)}}{\sqrt{\lamO\lamsO}},\\
  \cos\theta_{12} &= \frac{\sigma_3(\sigma_2-\sigma_1)}{\sqrt{\lamT\lamsT}},&
  \sin\theta_{12} &= \frac{2\sqrt{\sigma_3}\sqrt{\phi(s,\sigma_1,\sigma_3)}}{\sqrt{\lamT\lamsT}},\\
  \cos\hattheta &= \frac{2s(2m_\pi^2-\sigma_2)+(s+m_\pi^2-\sigma_1)(s+m_\pi^2-\sigma_3)}{\sqrt{\lamsO\lamsT}},&
  \sin\hattheta &= \frac{2\sqrt{s}\sqrt{\phi(s,\sigma_1,\sigma_3)}}{\sqrt{\lamsO\lamsT}}.
\end{align}
where we introduced the Kibble function $\phi$ as it enters all $\sin\theta$ expressions~\cite{Kibble:1960zz},
\begin{equation}
  \phi(s,\sigma_1,\sigma_3) = \sigma_1 \sigma_2 \sigma_3 - m_\pi^2 (s - m_\pi^2)^2,\quad  \sigma_2 = s + 3 m_\pi^2 - \sigma_1 - \sigma_3.
\end{equation}
We combined the $d$-functions in Eq.~\eqref{eq:rhoInt.through.ddd} and get the expressions for the angular part through invariants~\cite{10.7717/peerj-cs.103}:
\begin{equation} \label{eq:cos.angles}
  \sum_{\lambda,\lambda'}
  d^1_{\lambda0}(\theta_{23}) d^1_{\lambda\lambda'}(\hattheta) d^1_{\lambda'0}(\theta_{12}) =
  \cos (\theta_{12}+\hattheta-\theta_{23}) = \frac{\Hpoly}{\lambda^{1/2}_{1}\lambda^{1/2}_{3}\lamsO\lamsT},
\end{equation}
where $\Hpoly$ is a polynomial in $\sqrt{\sigma_1}$, $\sqrt{\sigma_3}$, and $\sqrt{s}$.
The expression \Hpoly is further factorized~\cite{Mathematica} and cancels terms
zeros of the denominator which otherwise would be pole singularities in the physical reason.
\begin{align}
  H(s,\sigma_1,\sigma_3) = &\sqrt{\sigma_1\sigma_3}\\\nonumber
  &\quad\times(\sqrt{s}-\sqrt{\sigma_1}-m_\pi)(\sqrt{s}-\sqrt{\sigma_1}+m_\pi)\\\nonumber
  &\quad\times(\sqrt{s}-\sqrt{\sigma_3}-m_\pi)(\sqrt{s}-\sqrt{\sigma_3}+m_\pi)\\\nonumber
  &\quad\times\Wpoly,
\end{align}
with the polynomial \Wpoly given by
\begin{align} \label{eq:Wpoly}
  \WpolyOpt{a}{b}{c} = &-4 m_\pi^6 + 4 m_\pi^2 s^2 - 4 m_\pi^4 a b +
   4 m_\pi^2 a^{3} b - 4 m_\pi^4 a c \\\nonumber
   &\quad +
   4 m_\pi^2 a^{3} c - 9 m_\pi^4 b c +
   8 m_\pi^2 a^2 b c + a^4 b c \\\nonumber
   &\quad+
   14 m_\pi^2 a b^2 c + 2 a^{3} b^2 c +
   9 m_\pi^2 b^{3} c - a^2 b^{3} c \\\nonumber
   & \quad -
   4 a b^4 c - 2 b^{5} c +
   14 m_\pi^2 a b c^2 + 2 a^{3} b c^2 + 12 m_\pi^2 b^2 c^2 \\\nonumber
   &\quad-
   6 a b^{3} c^2 - 4 b^4 c^2 + 9 m_\pi^2 b c^{3} -
   a^2 b c^{3} - 6 a b^2 c^{3} - 5 b^{3} c^{3} \\\nonumber
   &\quad-
   4 a b c^4 - 4 b^2 c^4 - 2 b c^{5}.
\end{align}
The angular function from Eq.~\eqref{eq:cos.angles} is simplified to its final form
\begin{equation}
  \cos (\theta_{12}+\hattheta-\theta_{23}) =
  \frac{\Wpoly}{((\sqrt{s}+\sqrt{\sigma_1})^2-m_\pi^2)((\sqrt{s}+\sqrt{\sigma_3})^2-m_\pi^2)\sqrt{(\sigma_1-4m_\pi^2)(\sigma_3-4m_\pi^2)}}.
\end{equation}
The final expression for the interference term is
\begin{align}\label{eq:rhoInt.final}
  \rhoINT(s) &= \frac{1}{2\pi(8\pi)^2 s} \int_{4m_\pi^2}^{\sPTH} \diff \sigma_1
\int_{\sigma_3^-(\sigma_1,s)}^{\sigma_3^+(\sigma_1,s)} \diff \sigma_3
\frac{\frho^*(\sigma_1)}{\sqrt{\sigma_1-4m_\pi^2}}\,
\frac{\frho(\sigma_3)}{\sqrt{\sigma_3-4m_\pi^2}}\\
&\qquad\times
\frac{\Wpoly}{((\sqrt{s}+\sqrt{\sigma_1})^2-m_\pi^2)((\sqrt{s}+\sqrt{\sigma_3})^2-m_\pi^2)}.
\end{align}

\section{The Dalitz plot integral in the complex plane}
\label{app:circling}
\begin{figure*}
  \includegraphics[width=0.8\textwidth]{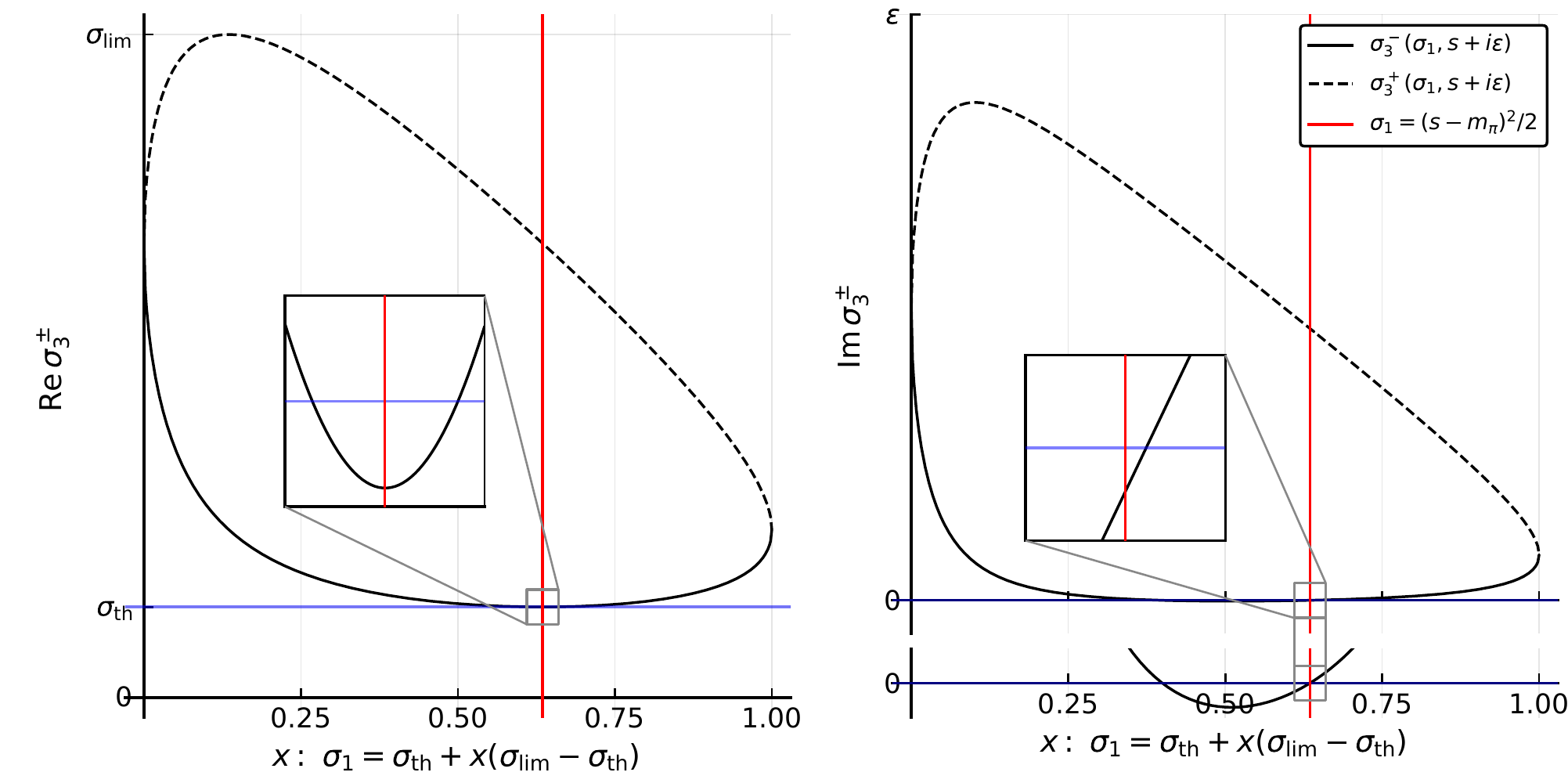}
  \caption{The left (right) plot presents the real (imaginary) part of the $\sigma_3^\pm$ as a function of $\sigma_1$ for a fixed value of $s+i\epsilon$.
  The $\sigma_1$ is changed linearly between the integration limits.
  The zoomed plots show how the $\sigma_3^-$ passes the real axis first below the branch point $\sTH = 4m_\pi^2$, then returns above the branch point performing the circling.
  The red line indicates the closest point on the $\sigma_1$-path to the $(s-1)/2$ since it does not go exactly through it.}
  \label{fig:sigma.pm.for.epsilon}
\end{figure*}

\begin{figure*}
  \includegraphics[width=0.45\textwidth]{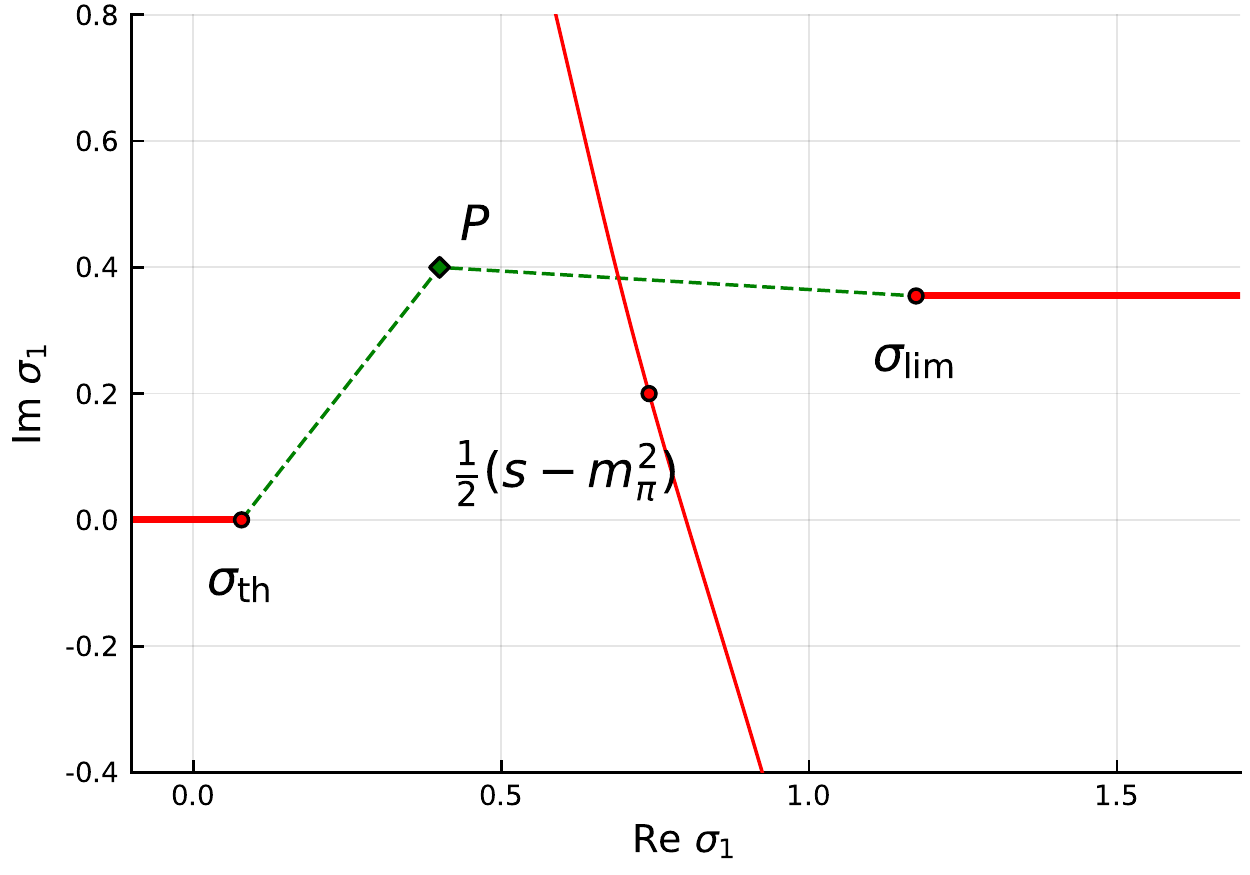}
  \caption{Analytical structure of the integrand in Eq.~\eqref{eq:remaining.sigma1.int}. The branch points are shown by the red dots with the cuts indicated by the solid red lines.
  The arbitrary integration path $\sTH\to\mathrm{P}\to\sLIM$ is shown by the dashed green line.}
  \label{fig:wall}
\end{figure*}

To address the issues of the evaluation of Eq.~\eqref{eq:rhoInt.final} for complex values of $s$,
we consider a simplified version of the problem:
\begin{equation}
  X(s) = \int_{\sTH}^{\sLIM}\diff\sigma_1 \int_{\sigma_3^-(\sigma_1,s)}^{\sigma_3^+(\sigma_1,s)} \diff\sigma_3\,\frac{1}{\sqrt{\sigma_3-4m_\pi^2}},
\end{equation}
where $\sTH = 4m_\pi^2$, $\sLIM = (\sqrt{s}-m_\pi)^2$, $\sigma_3^\pm(\sigma_1,s) = (s+3m_\pi^2-\sigma_1)/2 \pm \lamO^{1/2}\lamsO^{1/2}/(2\sigma_1)$.
Similar to Eq.~\eqref{eq:rhoInt.final} this expression contains two nested integrals with the same limits.
The integrand has a branch point at $4m_\pi^2$, the integration paths have to be modified in order to avoid crossing the cut.
The position of the $\sigma_3^\pm$ are shown in Fig.~\ref{fig:sigma.pm.for.epsilon} for $s = \re s + i\epsilon$.
We observe that the $\sigma_3^{+}$ has always positive imaginary part and stays far from the branch point $\sigma_3=4m_\pi^2$.
The $\sigma_3^{-}$ circles around the branch point changing the sheet of the integrand.
When $\sigma_1=\sLIM$, the $\sigma_3$ endpoints nearly coincide, $\sigma_3^{\pm}(\sLIM) = m_\pi(\sqrt{s}+m_\pi) \pm i\epsilon$, however,
they are on the different sides of the integrand cut.
In other words, if a straight line integration in $\sigma_3$ is done, we should observe a singularity related to the circling in the complex $\sigma_1$ plane.
The inner integral can be solved analytically.
\begin{equation} \label{eq:remaining.sigma1.int}
  X(s) = 2\int_{\sTH}^{\sLIM}\diff\sigma_1 \left(
    \sqrt{\sigma_3^+(\sigma_1,s)-4m_\pi^2}-
    \sqrt{\sigma_3^-(\sigma_1,s)-4m_\pi^2}
  \right),
\end{equation}
where the first term does not give problems near the physical region since $\sigma_3^+$ stays away from $4m_\pi^2$.
However, the second square root has two branch points at $(s-m_\pi^2)/2$ in the $\sigma_1$ plane.
(Another example of a simple function with two adjoined square root branch points is $\sqrt{z^2}$.)
\begin{equation}
  \sigma_3^+(\sigma_1,s)-4m_\pi^2 = 0 \quad \rightarrow \quad (\sigma_1-(s-m_\pi^2)/2)^2 = 0
\end{equation}
Fig.~\ref{fig:wall} shows the $\sigma_1$ plane, where we see that a straight connection between $\sTH$ and $\sLIM$ is not allowed by the presence of the cut.
Here, two options arise:
\begin{enumerate}
  \item $X^{(1)}$: we draw the $\sigma_1$ path directly through the branch point $(s-m_\pi^2)/2$
  (the point $\mathrm{P}$ in Fig.~\ref{fig:sigma.pm.for.epsilon} can be aligned with the branch point $(s-m_\pi^2)/2$).
  The point $\sigma_1 = (s-m_\pi^2)/2$ is special because when the path goes through it,
  the $\sigma_3^-$ does not circle the branch point but just touches it.
  \item $X^{(2)}$: we go analytically under the cut taking any arbitrary path.
  $X^{(2)}$ corresponds to the function which we would obtain connecting the points $\sigma_3^\pm$ properly, \ie avoiding $1/\sqrt{\sigma_3-4m_\pi^2}$ cut.
\end{enumerate}
The two options give two different analytical functions.
Additional discussions on the subject can be found in Ref.~\cite{Aitchison:1966lpz, Pasquier:1968zz}.

\bibliographystyle{apsrev4-1}
\bibliography{compass,hadron}
\end{document}